%% file: aistats_main.tex
\documentclass[twoside]{article}

\usepackage[accepted]{aistats2023}
\usepackage[utf8]{inputenc} 
\usepackage[T1]{fontenc}    
\usepackage[hidelinks]{hyperref}       
\usepackage{url}            
\usepackage{booktabs}       
\usepackage{amsfonts}       
\usepackage{nicefrac}       
\usepackage{microtype}      
\usepackage{graphicx}
\usepackage{float}
\usepackage{placeins}
\usepackage{listings}
\usepackage[ruled,vlined]{algorithm2e}
\usepackage{todonotes}
\usepackage{mathtools}
\usepackage{amsmath, amssymb, amsthm}
\usepackage{dsfont}
\usepackage{subfigure}
\usepackage{cleveref}
\usepackage{wrapfig}

\newtheorem{theorem}[]{Theorem}
\newtheorem{corollary}[]{Corollary}
\newtheorem{assumption}[]{Assumption}

\newtheorem{lemma}[]{Lemma}
\newcommand{\norm}[1]{\left\lVert#1\right\rVert}
\newcommand{\R}{\mathbb{R}}
\newcommand{\E}{\mathbb{E}}
\newcommand{\X}{\mathcal{X}}
\newcommand{\Z}{\mathcal{Z}}
\newcommand{\Fn}{\tilde{F}}
\newcommand{\C}{\mathcal{C}}
\newcommand{\x}{\pmb{x}}
\newcommand{\y}{\pmb{y}}
\newcommand{\z}{\pmb{z}}
\newcommand{\vb}{\pmb{v}}
\newcommand{\g}{\pmb{g}}
\newcommand{\gb}{\bar{\pmb{g}}}
\newcommand{\gh}{\hat{\pmb{g}}}
\newcommand{\Zb}{\pmb{Z}}

\begin{document}

%

%

\twocolumn[

\aistatstitle{High Probability Bounds for Stochastic Continuous Submodular Maximization}

\aistatsauthor{ Evan Becker \And Jingdong Gao \And  Ted Zadouri \And Baharan Mirzasoleiman}

\aistatsaddress{ Department of Computer Science, University of California Los Angeles (UCLA)} ]

\input{abstract}
\input{introduction}
\input{related_new}
\input{problem}
\input{results_new}

\input{experiments}

\input{conclusion}

\vspace{-2mm}\section*{Acknowledgment}\vspace{-3mm}
BM was supported the National Science Foundation CAREER Award 2146492.

\bibliography{ref}
\bibliographystyle{plain}

\appendix
\onecolumn
\input{appendix}

\end{document}

%% file: abstract.tex
\begin{abstract}
We consider maximization of stochastic monotone continuous submodular functions (CSF) with a diminishing return property. 
Existing algorithms only guarantee the performance \textit{in expectation}, and do not bound the probability of getting a bad solution. 
This implies that for a particular run of the algorithms, the solution may be much worse than the provided guarantee in expectation.
In this paper, we first empirically verify that this is indeed the case. Then, we provide the first \textit{high-probability} analysis of the existing methods for stochastic CSF maximization, namely PGA, {boosted PGA}, SCG, and SCG++. 
Finally, we provide an improved high-probability bound for SCG, under slightly stronger  assumptions,
with a better convergence rate than that of the expected solution.
Through extensive experiments on non-concave quadratic
programming (NQP) and optimal budget allocation,
we confirm the validity of our bounds and
show that even in the worst-case, PGA converges
to $OPT/2$, and boosted PGA, SCG, SCG++ converge to
$(1 - 1/e)OPT$, but at a slower rate than that of the expected solution.

\end{abstract}

%% file: introduction.tex
\section{INTRODUCTION}
While in general set functions are hard to optimize over, the subclass of submodular functions have useful properties that allow us to predictably achieve a certain approximation of the true optimal value in polynomial time \cite{wolsey1982analysis}. Submodular functions exhibit a natural diminishing returns property and appear in a wide variety of applications such as sensor placement \cite{10.1145/1127777.1127782}, graph cuts \cite{Jegelka2011SubmodularityBS}, data summarization \cite{Lin2011ACO}, marketing \cite{Kempe2003Maximizing} and clustering \cite{NIPS2005_b0bef4c9}. Thus, theoretical bounds on what optimization methods can achieve have important real-world implications. Continuous submodular functions (CSF) extend the notion of submodularity to continuous domains and provide an interesting class of non-convex functions that are still tractable to optimize over \cite{bian2017guaranteed}. 
CSFs have several applications, including non-convex/non-concave quadratic programming \cite{bian2017guaranteed}, robust budget allocation \cite{staib2017robust,soma2017non}, sensor energy management \cite{bian2017guaranteed}, online resource allocation \cite{eghbali2016designing}, learning assignments \cite{golovin2014online}, and e-commerce and advertising \cite{mehta2007adwords}.
In addition, they enable solving many discrete submodular problems efficiently through their
continuous relaxation such as multi-linear \cite{doi:10.1137/110839655} or Lovas extensions \cite{lovasz1982}. 
This has motivated a body of work on optimizing CSFs 
\cite{bian2017continuous,bian2017guaranteed,chekuri2015multiplicative}. 

More recently, constrained maximization of \textit{stochastic} submodular functions has gained a lot of attention \cite{Hassani2017gradient,karbasi2019stochastic,mokhtari2018conditional, zhang2022stochastic}. A stochastic CSF can be formulated as the expected value of 
stochastic functions $\Fn:\X\times\Z\rightarrow\R_+$:
\begin{equation}\label{eq:problem}
    \max_{\x\in \C} F(\x) = \max_{\x\in \C}\mathbb{E}_{\z\sim P}[ \Fn(\x, \z)],
\end{equation}
where $\C\subseteq \R^d_+$ is a bounded convex set,
$\x \!\in\! \X$ is the optimization variable, 
and $\z\!\in\!\Z$ is a random variable drawn from a (potentially unknown) distribution $P\!$.
Note that Problem \eqref{eq:problem} only assumes that $F(\x)$ is DR-submodular, and not necessarily the stochastic functions $\Fn(\x, \z)$. 
The continuous greedy algorithm \cite{bian2017guaranteed} can produce arbitrarily bad
solutions for Problem \eqref{eq:problem}, due to the non-vanishing variance of gradient approximations \cite{Hassani2017gradient}.
To address this, Projected Gradient Ascent (PGA) with diminishing step-sizes is first 
shown to provide a $[OPT/2\!-\!\epsilon]$ guarantee \cite{Hassani2017gradient}.
Later, Stochastic Continuous Greedy (SCG) suggested to reduce the noise of gradient approximations via a momentum term and provided a tight $[(1\!-\!1/e)OPT\!-\epsilon]$ guarantee \cite{mokhtari2018conditional}.
This work was followed by Stochastic Continuous Greedy++ (SCG++), which improved the complexity of SCG \cite{karbasi2019stochastic}, by leveraging a variance reduction technique 
\cite{fang2018spider}. {Most recently, boosted PGA algorithm using a non-oblivious function is proposed \cite{zhang2022stochastic}, which also achieves a $[(1\!-\!1/e)OPT\!-\epsilon]$ approximation guarantee. }\looseness=-1

However, the above algorithms only guarantee the performance of the solution \textit{in expectation}.
This implies that it is indeed possible that for a particular run of the algorithms, the optimizer gets extremely unlucky with its gradient estimates, and return a solution that is drastically 
worse than the provided guarantee in expectation.
Indeed, as we confirm by our experiments, all the algorithms for stochastic CSF, namely PGA, {boosted PGA}, SCG, and SCG++, 
may have a very high variance in their returned solution, as the noise gets larger. Crucially, the provided expectation bounds do not provide much insight, besides perhaps a basic Markov inequality, into the probability of getting these bad solutions.
This is because expected guarantees 
rely on bounding the variance of the gradient estimation error, and cannot bound the total accumulated error required for deriving high probability bounds. \looseness=-1

In this paper, we address the above question by providing the first 
high probability analysis of the existing methods for stochastic CSF maximization.
High-probability bounds have been explored very recently for 
the most popular optimization methods, namely, SGD \cite{harvey2019simple}, and momentum SGD \cite{li2020high}.
But, deriving high-probability bounds
for submodular optimization has remained unaddressed.
Different than the analysis of the expectation bounds for stochastic CSF maximization algorithms \cite{Hassani2017gradient,karbasi2019stochastic,mokhtari2020stochastic}, our analysis leverages two different strategies to bound the distance between the algorithmic solution and the optimal value. The first strategy is using a martingale process to model functions of the gradient noise, allowing for the use of Azuma-Hoeffding inequality to provide high-probability bounds on the algorithmic solution. The second strategy is to use Chebyshev's inequality to bound
sum of squared errors in gradient estimators, with a high probability. 
Table \ref{tbl:summary} summarizes our results.\looseness=-1

Our contributions are as follows.  
We derive the first high-probability analysis for stochastic CSF methods (under the same assumptions used for their expectation bounds), and show that after $K$ queries to the stochastic gradient oracle:
\begin{itemize}
    \vspace{-2mm}\item 
    For Projected Gradient Ascent (PGA) \cite{Hassani2017gradient} {and Boosted PGA \cite{zhang2022stochastic},} the lower-bound on the average function value during a run of the algorithm converges at rate $O(\frac{1}{K^{1/2}})$. 
    
    \vspace{-1mm}\item For Stochastic Continuous Greedy (SCG) \cite{mokhtari2018conditional}, the lower-bound on the final solution
    converges at rate $\mathcal{O}(\!\frac{\delta}{K^{1/3}})$, where $\delta$ depends on the confidence threshold.
    
    \vspace{-1mm}\item For Stochastic Continuous Greedy++ (SCG++)  \cite{karbasi2019stochastic}, the lower-bound on the final solution converges at rate 
    $\!\mathcal{O}(\!\frac{\delta}{K^{1/4}})$, where $\delta$ depends on the confidence threshold.

    \vspace{-1mm}\item 
    Under the sub-Gaussian assumption on the stochastic gradient oracle, we derive an improved high-probability bound on the final solution of SCG that converges to $(1-\frac{1}{e})OPT$ at a faster $\mathcal{O}(\frac{1}{K^{1/2}})$ rate. Interestingly, this rate even exceeds the 
    rate of convergence to the
    \emph{expected} solution provided by \cite{mokhtari2018conditional}.
    Our analysis involves providing the first high probability bound for adaptive momentum optimization methods, which can be  applied to other smooth function classes to provide superior convergence and generalization properties \cite{sun2021training}. Hence, it is of independent interest.
    
    \vspace{-1mm}\item Our extensive experiments on a non-concave quadratic programming example (NQP) and a realistic optimal budget allocation problem
    confirm the validity of our bounds and show that even in the worst-case PGA still converges to the $OPT/2$, 
    and  
    boosted PGA,
    SCG, SCG++ still converge to $(1\!-\!1/e)OPT$, 
    but at a slower rate. \looseness=-1
\end{itemize}
\vspace{-2mm}
Our results characterize the full distribution of the solutions for stochastic CSF maximization methods. In doing so, they allow an algorithm designer to answer questions about worst and best-case performance and even make modifications to mitigate the risk of getting a {bad solution}. 

\begin{table*}[t]
\centering
\caption{Comparison of existing expectation and our high-probability bounds for three stochastic monotone DR-submodular maximization algorithms, namely PGA, SCG, SCG++. Here $k$ is the number of queries to the stochastic gradient oracle $\nabla \Fn$. Note that while our original bound is not tight for SCG, by using the slightly stronger condition of a sub-Gaussian gradient noise one can achieve an $\mathcal{O}(1/K^{1/2})$ bound (see Sec. \ref{sec:improve_bounds}).}
    \begin{tabular}{l|lll}
    \toprule
    \textbf{Algorithm} & \textbf{Expectation Bound}                            & \textbf{Original Noise Assumptions}                                                      & \textbf{Bound Converges w.h.p?} \\ \midrule
    PGA \cite{Hassani2017gradient}                & $(\frac{1}{2})OPT - \mathcal{O}(\frac{1}{K^{1/2}})$   & $\nabla \tilde{F}$ bounded                                                        & Yes, at $\mathcal{O}(1/K^{1/2})$ rate                           \\
    {Boosted PGA \cite{zhang2022stochastic} }               & $(1-\frac{1}{e})OPT - \mathcal{O}(\frac{1}{K^{1/2}})$   & $\nabla \tilde{F}$ bounded                                                        & Yes, at $\mathcal{O}(1/K^{1/2})$ rate                           \\
    SCG \cite{mokhtari2018conditional}                & $(1-\frac{1}{e})OPT - \mathcal{O}(\frac{1}{K^{1/3}})$ & $Var(\nabla \tilde{F})$ bounded, sub-Gaussian*                                                   & Yes*, at $\mathcal{O}(1/K^{1/2})$ rate                                                              \\
    SCG++ \cite{karbasi2019stochastic}              & $(1-\frac{1}{e})OPT - \mathcal{O}(\frac{1}{K^{1/2}})$ & $\tilde{F}, \nabla \tilde{F}, \nabla^2\tilde{F}, \log(p(\z))$ bounded & Yes, at $\mathcal{O}(1/K^{1/4})$ rate \\\bottomrule
    \end{tabular}
\label{tbl:summary}
\end{table*}

%% file: related_new.tex
\section{RELATED WORK}
\vspace{-2mm}
\paragraph{Deterministic continuous submodular maximization.}
Maximizing deterministic continuous submodular functions have been first studied by Wolsey \cite{wolsey1982analysis}. More recently, \cite{chekuri2015multiplicative} proposed a multiplicative weight update algorithm that achieves $(1\!-\!1/e\!-\!\epsilon)$ approximation guarantee after $\tilde{O}(n/\epsilon^2)$ oracle calls to gradients of a monotone smooth twice differentiable DR-submodular function, subject to a polytope constraint ($n$ is the ground set size). Later, a conditional gradient method similar to the continuous greedy algorithm is shown to obtain a similar approximation factor after $O(n/\epsilon)$ oracle calls to gradients of monotone DR-submodular functions subject to a down-closed convex body \cite{bian2017guaranteed}. Such methods, however, require exact computation of the gradient of the function, which is not provided in the stochastic setting.

\vspace{-2mm}
\paragraph{Stochastic continuous submodular maximization.}
For stochastic continuous submodular maximization, conditional gradient methods 
may lead to arbitrarily poor solutions, due to the high noise variance \cite{Hassani2017gradient}.
While the noise variance can be reduced by averaging the gradient over a (large) mini-batch of samples at each iteration, averaging considerably increases the computational complexity of each iteration and becomes prohibitive in many applications.
To address this, stochastic proximal gradient method are first proposed. In particular, \cite{Hassani2017gradient}
showed that when the expected function is monotone and DR-submodular, Projected Gradient Ascent (PGA) provides a $OPT/2-\epsilon$ guarantee in $\mathcal{O}(1/\epsilon^2)$ iterations.
Later, \cite{mokhtari2018conditional} introduced Stochastic Continuous Greedy (SCG), which reduces the noise of gradient approximations via exponential averaging and achieves a $(1-1/e)OPT-\epsilon$ guarantee in expectation after $\mathcal{O}(1/\epsilon^3)$ iterations. 
More recently, Stochastic Continuous Greedy++ (SCG++)
improved the complexity of SCG to $\mathcal{O}(1/\epsilon^2)$, by using a stochastic path-integrated differential estimator (SPIDER) \cite{fang2018spider} to reduce the variance of the stochastic gradient.
Most recently, boosted PGA algorithm using a non-oblivious function is proposed \cite{zhang2022stochastic}, which also achieves a $[(1-1/e)OP T -\epsilon]$ approximation guarantee in $\mathcal{O}(1/\epsilon^2)$ iterations.
Existing works, however, only provide guarantees in expectation and cannot deliver any insight on the  distribution of the solutions or worst-case analysis. \looseness=-1

\vspace{-2mm}
\paragraph{High-probability bounds for stochastic submodular minimization.}\!\!\!\!
Very recently, \cite{zhang2021stochastic} studied an extension of the stochastic submodular \emph{minimization} problem, namely,
the stochastic $L^\natural$-convex \cite{murota1998discrete} minimization problem. $L^\natural$-convex functions are reduced to submodular functions when the ground set size is 2.
Specifically, \cite{zhang2021stochastic} developed a polynomial time algorithm that returns a near-optimal solution with a high probability. The proposed method relies on the Lovász extension of an $L^\natural$-convex function to transform the original problem to an equivalent continuous convex optimization problem, and applies the stochastic subgradient method to solve the continuous convex problem. 
To the best of our knowledge, high-probability bounds for continuous submodular \emph{maximization} have not been explored before.

%% file: problem.tex
\section{BACKGROUND AND PROBLEM FORMULATION}
\paragraph{Continuous submodular functions.}
We start by reviewing definition of submodularity for set functions.
A set function $f: 2^V \rightarrow \R_+$, defined on the ground set $V$ , is submodular if for all subsets $A, B \subseteq V$, we have that
\begin{equation}
f(A) + f(B) \geq f(A \cup B) + f(A \cap B). 
\end{equation}
The notion of submodularity can be extended to continuous domains. 
A continuous function $F: \X \rightarrow \R_+$ defined on the set $\X = \Pi_{i=1}^n \X_i$, where each $\X_i$ is a compact subset of $\R_+$, is continuous submodular if for all $\x, \y \in \X$ we have 
\begin{equation}
F(\x) + F(\y) \geq F(\x \vee \y) + F(\x \wedge \y).
\end{equation}
Here, $\x \vee \y := \max(\x, \y)$ is component-wise maximum and $\x \wedge \y := \min(\x, \y)$ is component-wise minimum operations. 
A submodular function $F$ is monotone on $\X$, if for every $\x, \y \in \X, \x \leq \y$ we have that $F(\x) \leq F(\y)$.
A function $F$ defined
over $\X$ satisfies the diminishing returns (DR) property, if for every $\x, \y \in \X, \x \leq \y$, and 
any standard basis
vector $e_i \in \R^n$ and any
$k \in \R_+$ s.t. $(ke_i + \x)\in\X$ and $(ke_i + \x)\in\X$, it holds that
\begin{equation}
    f(ke_i + \x) - f(\x) \geq f(ke_i + \y) - f(\y).
\end{equation}
When $F$ is twice-differentiable, 
DR-submodularity implies that all diagonal entries of the Hessian are non-positive \cite{bian2017guaranteed}. I.e.,\looseness=-1
\begin{equation}\label{eq:dr-hessian}
\forall i= j, \quad \forall \x \in\X  \quad \frac{\partial^2 F(\x)}{\partial x_i \partial x_j}\leq 0.
\end{equation}

\paragraph{Stochastic continuous submodular maximization.} 
In this work, we focus on constrained maximization of stochastic continuous DR-submodular functions. 
Formally, our goal is to find $\x^*$ that maximizes the expected value $F(\x)$ of the stochastic function $\Fn(\x,\z)$ over $\x$, where the expectation is with respect to the random variable $\Zb$: 
\begin{equation}
    \max_{\x\in\C} F(\x):= \max_{\x\in\C} \mathbb{E}_{\z\sim P}[\Fn(\x,\z)],
\end{equation}
where $\C \!\subseteq \!\R_+$ is a convex compact set, and $\z$ is the realization of the random variable $\Zb$ drawn from a distribution $P$. 
We assume that the expected objective function $F(\x)$ is monotone and DR-submodular and the stochastic functions $\Fn(\x, \z)$ may not be monotone nor submodular.
We denote by $OPT \!\triangleq \max_{x\in \mathcal{C}} F(\x)$ the optimal value of $F(\x)$ over $\C$. \looseness=-1 


%% file: results_new.tex
\section{HIGH-PROBABILITY BOUNDS FOR STOCHASTIC CONTINUOUS SUBMODULAR MAXIMIZATION}
Next, we discuss our high-probability bounds for stochastic CSF 
maximization algorithms, namely Projected Gradient Ascent (PGA), {boosted PGA,} Stochastic Continuous Greedy (SCG), and Continuous Greedy++ (SCG++).

\subsection{Projected Gradient Ascent}
We start by analyzing the worst-case performance of the PGA method 
which achieves a $[OPT/2-\epsilon]$ approximation in expectation, in $\mathcal{O}(1/\epsilon^2)$ iteration
\cite{Hassani2017gradient}.
PGA starts from an initial estimate $\x_0\in\mathcal{C}$. Then at every iteration $t$, it takes a step in the direction of the noisy gradient $\g_t = \nabla \tilde{F}(\x_t,\z_t)$, and projects the solution
onto the convex set $\mathcal{C}$. The update rule at step $t$ takes the following form:
\begin{equation}
    \x_{t+1}=\mathcal{P}_\C(\x_t + \mu_t \g_t),
\end{equation}
where, $\mu_t$ is the diminishing learning rate at step $t$, and $\mathcal{P}_\C$ denotes the Euclidean projection onto the set $\C$.
The pseudocode is provided in Appendix \ref{apx:alg}. 

Hassani et al. \cite{Hassani2017gradient}
provided a lower bound on the expected function value, $\E[F(\x_{\tau})]$, at
a time-step $\tau$, 
sampled uniformly at random from
$\{1,\dots, T\}$. 
Important to note, however, is that the derived expectation is not only over this random variable $\tau$, but \emph{also the noise} coming from the gradient estimates. This
implies that 
it is possible that the optimizer gets extremely unlucky with its gradient estimates, in which case no $\x_t$ 
satisfies the lower bound (for example, consider the unlikely but still possible scenario where $\forall t, \: \g_t=0$). In Theorem \ref{thm:pga_bound}, we provide an exact answer for how unlikely a failure event like this would be.\looseness=-1


To do so, we make similar assumptions to \cite{Hassani2017gradient}: 

\begin{assumption}\label{ass:x_bound}
\label{as:bounded}
The diameter of the constraint set $\C$ is bounded by $D$. I.e., $\forall \x, \y \in \C$, we have 
\begin{equation*}
    \norm{\x-\y}\leq D.
\end{equation*}
\end{assumption}
\begin{assumption}\label{ass:g_lip}
\label{as:smooth}
The function $F$ is Lipschitz smooth with constant $L$, over $\X$. I.e., $\forall \x, \y \in \C$, we have
\begin{equation*}
    \norm{\nabla F(\x) -\nabla F(\y)} \leq L\norm{\x-\y}.
\end{equation*}
\end{assumption}
\begin{assumption}\label{ass:g_bound}
Stochastic gradients $\g_t\!=\!\nabla \Fn(\x,\z)$ are bounded in distance from their mean $\nabla {F}(\x_t)=\E[\g_t]$:
\begin{equation*}
    \norm{\nabla {F}(x_t)-\g_t}\leq M.
\end{equation*}
\end{assumption}
%
The following theorem shows that for any fixed confidence interval $p$, the lower bound on $\sum_{t\in[T]}F(\x_t)/T$ will converge to $OPT/2$ at a rate of $\mathcal{O}(\sqrt{\frac{\log(1/1-p)}{T}})$.
\begin{theorem}\label{thm:pga_bound}
    Consider running PGA for $T$ iterations with step size of $\eta_t=\frac{2}{\sqrt{t}}$ with Assumptions \ref{ass:x_bound},\ref{ass:g_lip},\ref{ass:g_bound} satisfied. Then with probability $p\!\geq\! 1\!-\!\delta$, where $\delta \!\in\! [0,1]$, the average function value returned by the algorithm is lower bounded by\looseness=-1
    \begin{align}
        \frac{1}{T}\sum_{t=1}^T F(\x_t) &\geq \frac{1}{2}OPT  
         -\frac{C}{\sqrt{T}}
         - DM\sqrt{\frac{\log(1/\delta)}{2T}} \\ \nonumber
        &\geq \frac{1}{2}OPT -  \mathcal{O}\left(\sqrt{\frac{\log(1/\delta)}{T}}\right),
    \end{align}
    where we denote the constant $C:= \left(\frac{8(L+M)^2 +D^2}{8}\right)$.
\end{theorem}
Unlike the expectation bound provided in \cite{Hassani2017gradient}, Theorem \ref{thm:pga_bound} assures that with high probability, \emph{at least one iterate} from a \emph{single} algorithm run will be larger than the lower bound. Therefore, 
one could modify the default PGA algorithm to return the best iterate, 
$\max_{t\in[T]}F(\x_t)$, 
which is guaranteed to also be lower bounded with high probability by $OPT/2$.
We note that even in the case where the true function evaluation is hard to compute, one can still find the best iterate with high probability given unbiased stochastic function evaluations $\tilde{F}(\x,\z)$. Concretely, given the ordering of iterates from highest to lowest function value $\x_{[1]},\dots,\x_{[T]}$, consider the difference between the best two solutions $d:=F(\x_{[1]})-F(\x_{[2]})$. One can use a Hoeffding bound to determine the relatively small number of samples, $m$, needed to calculate $\bar{F}(\x):=\frac{1}{m}\sum \tilde{F}(\x)$ for each iterate, such that
$\bar{F}(\x_{[1]})\!>\! F(\x_{[1]})-d/2$ and $\bar{F}(\x_{[k]})\!<\! F(\x_{[2]})+d/2$ for all $k>1$ occurs with very high probability. 
Alternatively, since $F(\x)\leq OPT\:\: \forall \x$, 
at least 
$\frac{r}{r + (1/2)OPT}$ fraction of solutions are greater than $\sum_{t\in[T]}F(\x_t)/T-r$ for a slack variable $r$.
That is, with only $k$ true function evaluations, 
at least one good solution is found with probabiliy $p>1-(1-r)^k$. \looseness=-1 

Theorem \ref{thm:pga_bound} 
relies on diminishing returns and smoothness of $F$, along with the bound on $\C$ to 
first bound the difference between $F(\x_t)$ and $F(\x^*)$
based on 
the inner product between gradient noise and $\x_t-\x^*$.
However, instead of taking the expectation of this inequality, 
it directly shows that with Assumption \ref{ass:g_bound}, these random products satisfy the conditions of a c-lipschitz Martingale difference sequence.
This allows using standard high probability bounds (Azuma-Hoeffding). 
See Appendix \ref{sec:pga_proof} for the full proof.


We note that Assumption \ref{ass:g_bound} is stronger than simply bounded variance and is necessary to apply the Azuma-Hoeffding inequality. However if $\g_t = \nabla F(\x_t) + \z_t$ with each $\z_t$ being zero mean and Sub-Gaussian, a similar version of Theorem \ref{thm:pga_bound} 
can be derived following  \cite{harvey2019simple}. 


\begin{corollary}\label{cor:pga_bound}
    Consider the case where we set $\delta = \exp(-\sqrt{T})$.
    Then the averaged function value of PGA is lower bounded with probability $p\geq 1-2\exp(-\sqrt{T})$ by:
    \begin{equation}
        \frac{1}{T}\sum_{t=1}^T F(\x_t) \geq \frac{1}{2}OPT -  \mathcal{O}\left(\frac{1}{T^{1/4}}\right).
    \end{equation}
\end{corollary}
We see that as $T\rightarrow \infty$ we have both $p\rightarrow 1$ and $\!\mathcal{O}(1/T^{1/4})\!\rightarrow 0$. Thus, our lower-bound is tight.
{
\subsection{Boosted Projected Gradient Ascent}
Very recently, boosted PGA 
\cite{zhang2022stochastic} 
is proposed to provide 
$[(1-\frac{1}{e}-\epsilon^2)OPT]$ approximation in expectation, in $\mathcal{O}(1/\epsilon^2)$ iterations.
The idea is to find an auxiliary (non-oblivious) function that can provide a better approximation guarantee than the original DR-submodular function. 
Then the stochastic gradients of the non-oblivious function $F'$ (instead of the stochastic gradient of the original DR-submodular function $F$) are leveraged by PGA.
Specifically, \cite{zhang2022stochastic} used 
the following non-oblivious function $F'$ and its gradient $\nabla F'$: \looseness=-1
\begin{align}
    F'(\x) := \int_0^1 \frac{e^{(s-1)}}{s}F(s*\x) ds, \\
    \nabla F'(\x) := \int_0^1 e^{(s-1)}\nabla F(s*\x) ds.
\end{align}
This has the nice property that $\langle \y-\x, \nabla F'(\x)\rangle \geq (1-1/e)F(\y)-F(\x)$, which guarantees 
$(1-1/e)OPT$ approximation. Additionally, when the original function is Lipschitz smooth (Assumption \ref{as:smooth}), $F'(\x)$ is 
Lipschitz smooth with constant $L' = L(1+1/e)$. To efficiently approximate $\nabla F'$ given a noisy gradient estimate $\nabla \tilde{F'}$, \cite{zhang2017empirical} uses the following estimator:
\begin{align}\label{eq:nonobv-noisy}
    \nabla \tilde{F}'(\x_t) := (1-\frac{1}{e})\nabla \tilde{F'}(s_t*\x_t).
\end{align}
Here, $s_t$ is independently sampled from a distribution 
$\mathbb{P}(\pmb{S}\!\!\leq\!\! s) \!=\!\! \int_0^s \frac{1}{1-e^{-1}} \mathds{1}(u\!\in\![0,1])du$, where $\mathds{1}$ is indicator function.\looseness=-1

The following theorem provides a lower bound on $\sum_{t\in[T]}F(\x_t)/T$, with high probability.
\begin{theorem}\label{thm:nonobv_pga_bound}
  Consider running boosted PGA for $T$ iterations with step size of $\eta_t=\frac{2}{\sqrt{t}}$ with Assumptions \ref{ass:x_bound},\ref{ass:g_lip},\ref{ass:g_bound} satisfied. Then with probability $p\!\geq\! 1\!-\!\delta$, where $\delta \!\in\! [0,1]$, the average function value returned by the algorithm is lower bounded by\looseness=-1
    \begin{align}
        \frac{1}{T}\sum_{t=1}^T F(\x_t) &\geq (1-\frac{1}{e})OPT  
         -\frac{C'}{\sqrt{T}}
         - DM'\sqrt{\frac{\log(1/\delta)}{2T}} \nonumber\\ 
        &\geq (1-\frac{1}{e})OPT -  \mathcal{O}\!\left(\!\sqrt{\frac{\log(1/\delta)}{T}}\right),
    \end{align}
    where we denote the constant $C':= \left(\frac{8(L' D+M')^2 +D^2}{8}\right)$, and constant $M':= (M+2 LD)\left(1-\frac{1}{e}\right)$. 
\end{theorem}
Note that this bound has the same rate of convergence as \Cref{thm:pga_bound} up to a constant factor. This similarity also means a result parallel to \Cref{cor:pga_bound} can be derived, demonstrating that this algorithm will also converge with $p\rightarrow 1$ as $T\rightarrow \infty$.  
The proof of \Cref{thm:nonobv_pga_bound} follows a similar structure to \Cref{thm:pga_bound}. We bound the sum of differences $F(\x_t)-F(\x^*)$ by a Martingale sequence with a bounded difference property. The key distinction in the non-oblivious case is that we must bound gradients from $\nabla F'(\x)$ as well as $\nabla \tilde{F}'(\x)$.  We defer the full proof to Appendix \ref{apx:nonobv-proof}.\looseness=-1

\textbf{Guarantees for weakly submodular functions.} We note that Theorems \ref{thm:pga_bound}, \ref{thm:nonobv_pga_bound} for PGA and boosted PGA can be extended to $\gamma$-weakly DR-submodular functions, where we have 
$\gamma = \inf_{\x \leq \y}\inf_{i}([\nabla F(\x)]_i/[\nabla F(\y)]_i)$). This setting produces bounds with the same rate of convergence up to a constant, to $(\frac{\gamma^2}{1+\gamma^2})OPT$ and $(1\!-\!e^{-\gamma})OPT$ for PGA and Boosted PGA, respectively.
Note that $\gamma\!=\!1$ indicates a differentiable and monotone DR-submodular function.
\looseness=-1
}

\subsection{Stochastic Continuous Greedy}
Next, we analyze the worst-case performance of the Stochastic Continuous Greedy (SCG) algorithm. SCG uses a momentum term to reduce the noise of gradient approximations.
%
It starts from $\x_0=\pmb{0}$, and at every iteration $t$, 
calculates:
\begin{equation}
    \gb_{t+1}=(1-\rho_t)\gb_t+\rho_t \nabla \Fn(\x_t,\z_t),
\end{equation}
where $\rho_t$ is a stepsize which approaches zero as $t$ approaches infinity, and $\gb_0=0$. 
The SCG is then ascent in the 
direction: 
\begin{equation}
    \vb_t \gets \arg\max_{\vb \in \C}\{\left<\gb_{t}^{T}, \vb\right>\}, 
\end{equation}
using the following updates rule with step-size ${1}/{T}$:
\begin{equation}
    \x_{t+1}=\x_t+\frac{1}{T}\vb_t.
\end{equation}
The stepsize $\frac{1}{T}$ and the initialization $\x_0 = \pmb{0}$ ensure that after $T$ iterations the variable $\x_T$ ends up in the convex set $\C$. 
The pseudocode is provided in Appendix \ref{apx:alg}.

SCG provides a tight $[(1-1/e)OPT-\epsilon]$ guarantee in expectation for the last iterate $T$, with $\mathcal{O}(1/\epsilon^3)$ stochastic gradient computations \cite{mokhtari2018conditional}. 
But, 
similar to 
PGA \cite{Hassani2017gradient},
the expected 
guarantee of SCG does not tell us how frequently \textit{bad solutions}, with $F(\x_T)\!<\![(1-1/e)OPT-\epsilon]$ 
are returned.

Here, we answer the above question by providing a high-probability bound on the value of the final solution, $F(\x_T)$, returned by SCG. 
To do so, instead of 
assuming bounded gradient error
(Assumption \ref{ass:g_bound}), we use the weaker assumption from \cite{mokhtari2018conditional} 
on the variance of the stochastic gradients: 
\begin{assumption}\label{ass:g_var}
Stochastic gradients have mean $\E[\g_t]= \nabla {F}(\x_t)$ and bounded variance:
\begin{equation*}
    \E_{z\sim p}\left[\norm{\g_t-\nabla F(\x)}^2\right] \leq \sigma^2. 
\end{equation*}
\end{assumption}
Given Assumption 4, 
\cite{mokhtari2018conditional} showed that the variance of the momentum error, i.e., $\E[\norm{F(\x_t) - \gb_t}^2]$, converges to zero, 
as $t$ grows. 
{
However, this cannot be directly used to provide a high-probability bound on the value of the final solution (as we require the summation rather than the expectation of error terms).}
To address this, instead of using the bound on the variance of the noisy gradient at step $t$, 
we apply Chebyshev's inequality to bound the probability of the noisy gradient to be far away from its expectation. Then, we 
use a union bound on iterations 
$t\!\in\![T\!-\!1]$
to 
get the
next Lemma: \looseness=-1
\begin{lemma}\label{lemma:scg}
    Consider the Stochastic Continuous Greedy algorithm, with
    $\rho_t = \frac{4}{(t+8)^{2/3}}$. 
    Under Assumptions \ref{ass:x_bound}, \ref{ass:g_lip}, \ref{ass:g_var}, 
    we have the following high probability bound on the total variance of the noisy gradients during $t\in\{0,\cdots,T-1\}$:
       \begin{equation}
           \mathbb{P}\left(\sum_{t=0}^{T-1}\norm{\nabla F(\x_t) - \gb_t}^2 \leq \delta^2 \sum_{t=0}^{T-1}\frac{Q}{(t+9)^{2/3}}\right) \geq 1-\frac{T}{\delta^2},
       \end{equation}
    where $\!Q\!\!:=\!\!\max\! \left\{\!\norm{\nabla\! F(\x_0) \!-\! \gb_0}^2 \!9^{2/3}\!,\! 16\sigma^2\!\!+\!3L^2D^2\!\right\}$, \!$\delta\!\!>\!0$. 
\end{lemma}
The proof can be found in \Cref{apx:thm2}. Note that Lemma \ref{lemma:scg} does not rely on taking the expectation of the function values or gradient approximations. Equipped with Lemma \ref{lemma:scg}, we derive the following lower-bound on $F(\x_T)$, by recursively bounding the distance of the iterates to the optimum, $F(\x^*)$.
\begin{theorem}\label{thm:scg_bound}
    Consider the Stochastic Continuous Greedy algorithm, with $\rho_t = \frac{4}{(t+8)^{2/3}}$. 
    Under Assumptions \ref{ass:x_bound}, \ref{ass:g_lip}, \ref{ass:g_var}, we have that with probability greater than $1-\frac{T}{\delta^2}$:
    \begin{align}
    F(\x_T) &\geq (1-\frac{1}{e})F(\x^*) - \delta\frac{2Q^{1/2}D}{T^{1/3}} -\frac{LD^2}{2T^2} \nonumber\\
    &= (1-\frac{1}{e})OPT - \mathcal{O}(\frac{\delta}{T^{1/3}}),
    \end{align}
    where $\!Q\!\!:=\!\!\max\! \left\{\!\norm{\nabla\! F(\x_0) \!-\! \gb_0}^2 \!9^{2/3}\!,\! 16\sigma^2\!\!+\!3L^2D^2\!\right\}$, \!$\delta\!\!>\!0$. 
\end{theorem}
The proof can be found in Appendix \ref{apx:thm2}. Unlike our Theorem \ref{thm:pga_bound} for PGA, we see that for any fixed confidence threshold $p$ implying $\delta = \sqrt{\frac{T}{1-p}}$, our lower bound does not converge. This is a direct result of the weakening of the noise assumption, since the gradient noise may have bounded variance but not be bounded itself. Next, we provide an improved bound when gradient noise is sub-Gaussian.
 

\subsubsection{Improved Bound under Sub-Gaussian Noise}\label{sec:improve_bounds}
The weak assumptions on the noisy gradients (only bounded variance from Assumption \ref{ass:g_var}) make it difficult to apply typical Martingale or sub-Gaussian inequalities. If instead, we assume that the noisy gradient approximations are sub-Gaussian, we arrive at a surprisingly tight lower bound. First we describe the sub-Gaussian noise assumption as follows:
\begin{assumption}\label{ass:subG}
Stochastic gradients have mean $\E[\g_t]= \nabla {F}(\x_t)$ and $\hat{\z}_t:=\norm{\g_t\!-\!\nabla F(\x)}$ is sub-Gaussian. I.e. for $\sigma>0$, we have: 
\begin{equation}
    \mathbb{E}(e^{\lambda\hat{\z}_t^2})\leq e^{\sigma^2\lambda^2/2} \quad \forall\lambda \in \R 
\end{equation}
\end{assumption}
When using the SCG algorithm under this new assumption, we can derive the following high probability bound:

\begin{theorem}\label{thm:scg_strong}
Consider the Stochastic Continuous Greedy algorithm, with $\rho_t = \frac{1}{t^\alpha}$ where $\alpha\in (0,1)$. Then under Assumptions \ref{ass:x_bound}, \ref{ass:g_lip}, \ref{ass:subG}, 
with probability greater than $1-\delta$:
\begin{align}
        F(\x_T) &\geq (1-\frac{1}{e})OPT - \frac{2DK\sigma\sqrt{\log(1/\delta)}}{T^{1/2}}\\
        &\quad\quad\quad\quad\quad\quad\quad\quad- (\frac{4K+1}{2})\frac{LD^2}{T} \nonumber \\
        &= (1-\frac{1}{e})OPT - \mathcal{O}\left(\frac{\sqrt{\log(1/\delta)}}{T^{1/2}}\right),
\end{align}
where $K:=\frac{1}{1-\alpha}\Gamma\left(\frac{1}{1-\alpha}\right)$.
\end{theorem}
At a high level, the proof of Theorem \ref{thm:scg_strong}
expands the momentum into a weighted summation of gradient approximations, which 
after some careful manipulations 
can be treated as a summation of sub-Gaussian variables. 
This is to our knowledge the first such result for adaptive momentum optimization methods, where the momentum can change over time. Notably, it is general enough to be used even in the context of other
smooth function classes. Adaptive momentum enjoys some superior convergence and generalization
properties \cite{sun2021training}.
See \cref{sec:scg_strong} for the detailed proof.

Notably, the bound in Theorem \ref{thm:scg_strong} has a faster convergence rate than the original expectation bound of \cite{mokhtari2018conditional}, i.e., $(1-\frac{1}{e})OPT-\mathcal{O}(1/T^{1/3})$. The new bound suggests that for certain well conditioned problems, SCG can achieve the same convergence rate as SCG++. 
In our experiments in Sec. \ref{sec:experiments}, we show that empirically the distribution of solutions of SCG do converge when gradient noise is (sub-)Gaussian.

\subsection{Stochastic Continuous Greedy++}
Finally, we analyze the worst-case performance of the Stochastic Continuous Greedy++ (SCG++),
which aims to speed up SCG, using a stochastic path-integrated differential estimator (SPIDER) \cite{fang2018spider} for the gradient. 
SCG++ assumes that  the probability distribution of the random variable $\z$ depends on the variable $\x$ and may change during the optimization. 
To obtain an unbiased gradient estimator $\gh_t$ with a reduced variance, SCG++ uses a mini-batch of 
samples to first get an unbiased estimate of 
the Hessian, $\tilde{\nabla}^2_t$:
\begin{equation}
    \tilde{\nabla}^2_t=\frac{1}{|\mathcal{M}|}\sum_{(a,\z(a))\in\mathcal{M}}\tilde{\nabla}^2 F(\x(a),\z(a)),
\end{equation}
where $a$ is selected uniformly at random from $[0,1]$, $\z(a)$ is a random variable with probability distribution $p(\z(a); \x(a))$, $\x(a) := a\x_t + (1-a)\x_{t-1}$, and $\mathcal{M}$ is a mini-batch containing $|\mathcal{M}|$ samples of random tuple $(a, \z(a))$.
Then, SCG++ uses the Hessian estimate to recursively calculate unbiased estimates of the gradient, based on the gradient differences  $\tilde{\Delta}^t$:
\begin{equation}
    \tilde{\Delta}^t:=\tilde{\nabla}^2_t(\x_t-\x_{t-1}). 
\end{equation}
A gradient estimate, $\gh_t$, with a reduced variance is then calculated as the initial noisy gradient estimate plus the sum of all the gradient differences 
up to time $t$: \looseness=-1 
\begin{align}
    \gh_t&=\nabla\Fn(\x_0,\mathcal{M}_0)+\sum_{i=1}^i\tilde{\Delta}^t.
\end{align}
With the above gradient estimate $\gh_t$, SCG++ starts from $\x_0=\pmb{0}$, and at each iteration $t$, performs a standard Frank-Wolfe step with step-size $\frac{1}{T}$. 
The full update sequence is provided in Appendix \ref{apx:alg}.

SCG++ converges in expectation to the same $[(1-1/e)OPT-\epsilon]$ approximation as SCG, but using only $\mathcal{O}(1/\epsilon^2)$ stochastic gradient evaluations and $\mathcal{O}(1/\epsilon)$ calls to the linear optimization oracle \cite{karbasi2019stochastic}. However, similar to PGA and SCG, the expected approximation guarantee of SCG++ does not show the probability of returning a final solution that is much lower than the expected solution.

The analysis of SCG++ requires stronger assumptions than SCG. Besides the 
monotone DR-submodularity of $F$ and the bounded diameter on $\C$  (Assumption \ref{ass:x_bound}), we use the same assumptions originally used to analyze SCG++ in \cite{karbasi2019stochastic}:
\begin{assumption}
The function value at the origin is:
\begin{equation}
    F(\mathbf{0}) \geq 0.
\end{equation}
\end{assumption}
\begin{assumption}
The stochastic function $\Fn(\x,\z)$, its gradient, and its Hessian are bounded:
\begin{align}
    \Fn(\x,\z) &\leq B, \\
    \norm{\nabla \Fn(\x,\z)} &\leq G_{\Fn}, \\
    \norm{\nabla^2 \Fn(\x,\z)} &\leq L_{\tilde{F}}.
\end{align}
\end{assumption}

\begin{assumption}
The function $\log p(\z)$ has the following bounds on gradient and Hessian:
\begin{align}
    \mathbb{E}\left(\norm{\nabla \log p(\z)}^4\right) \leq G_p^4, \\
    \mathbb{E}\left(\norm{\nabla^2 \log p(\z)}^2\right) \leq L_p^2.
\end{align}
\end{assumption}

\begin{assumption}
The Hessian of the stochastic function $\Fn(\x,\z)$ is $L_2$ Lipschitz continuous with constant $L_{2,{\Fn}}$. The Hessian of the function $\log p(\z)$ is $L_2$ Lipschitz continuous with constant $L_{2,p}$.
\begin{equation}
    \norm{\nabla^2\Fn(\x,\z)-\nabla^2\Fn(\y,\z)} \leq L_{2,{\Fn}} \norm{\x-\y},
\end{equation}
\begin{equation}
    \norm{\nabla^2\log p(\z)-\nabla^2\log p(\z)} \leq L_{2,p} \norm{\x-\y}.
\end{equation}
\end{assumption}

Under the above assumptions and with $\mathcal{O}(\epsilon^{-1})$ calls to the stochastic oracle per iteration, the variance of the gradient approximation $\gh_t$ converges to zero 
\cite{karbasi2019stochastic}.
Instead of directly upper bounding variance as in \cite{karbasi2019stochastic},
we use Chebyshev's inequality to prevent any expectations from appearing in our bound. Specifically we use the following Lemma: 
\begin{lemma}\label{lemma:scgplus}
    Given SCG++ under Assumptions 1,5-8, we have the following high probability bound: 
    \begin{equation}
    \mathbb{P}\left(\sum_{t=0}^{T-1}\norm{\nabla F(\x_t) - \gh_t}^2 \leq \delta^2\sum_{t=0}^{T-1}\frac{2L^2D^2}{t^2}\right) \leq 1-\frac{T}{\delta^2}.
    \end{equation}
\end{lemma}
Lemma \ref{lemma:scgplus} allows us to directly bound the function value of last iterate of SCG++, $F(\x_T)$, with a high probability.
\begin{theorem}
\label{thm:scg_plus}
    Consider applying SCG++ under Assumptions 1, 5-8. Then with probability $1-\frac{T}{\delta^2}$ :
    \begin{align}
        F(\x_T) &\geq (1-\frac{1}{e})F(\x^*) - \delta\frac{LD^2}{T^2} -\frac{LD^2}{2T^2} \nonumber\\
        &= (1-\frac{1}{e})OPT - \mathcal{O}(\frac{\delta}{T}).
    \end{align}
\end{theorem}

For a fixed probability threshold 
$p$ we get the next Corollary: \looseness=-1
\begin{corollary}\label{cor:scg_plus} For $\delta \!=\! \sqrt{\frac{T}{1-p}}$, with probability greater than $p$, we have: 
    \begin{align}
    F(\x_T) &\geq (1-\frac{1}{e})F(\x^*) - \frac{LD^2}{\sqrt{1-p}}\frac{1}{\sqrt{T}} -\frac{LD^2}{2T^2} \\
    &= (1-\frac{1}{e})OPT - \mathcal{O}(\frac{1}{\sqrt{T}})
\end{align}
\end{corollary}

SCG++ makes $\mathcal{O}(T)$ queries to the stochastic gradient oracle per iteration, and $K=T^2$ queries in total. Hence, with probability greater than $p$ the bound in Corollary \ref{cor:scg_plus} becomes:
\begin{equation}
    F(\x_T) \geq  (1-\frac{1}{e})OPT - \mathcal{O}(\frac{1}{K^{1/4}}).
\end{equation}
For any fixed confidence interval $p$, the lower bound still converges to $(1-\frac{1}{e})OPT$, albeit at a slower rate. 
However, we believe that a tighter $1/\sqrt{K}$ high probability bound likely exists, as evident by our experimental results in Sec. \ref{sec:experiments}.\looseness=-1

%% file: experiments.tex
\section{NUMERICAL RESULTS} \label{sec:experiments}

In our experiments, we first show that bad solutions of PGA, {boosted PGA} SCG, and SCG++ can be much worse than their expected values. Then, we validate our proposed bounds 
on simulated and real-world datasets. 
In practice, due to measurement errors or inexact function calculations,
the function and thus
the gradient evaluations are often noisy. In such situations, our high probability bounds
can effectively quantify the worst-case performance and be utilized to mitigate the risk of getting a bad solution.

\subsection{Continuous Submodular Problems}\label{sec:csp}
First, we introduce three monotone continuous DR-submodular problems that we use in our experiments.

\vspace{-2mm}
\paragraph{\!Non-convex/non-concave quadratic \!programming (NQP).}
NQP functions of the form $f(\x) = \frac{1}{2} \x^T \pmb{H} \x + \pmb{h}^T\x$ arise in many applications, including scheduling, inventory theory, and free boundary problems \cite{bian2017continuous}. When all off-diagonal entries of $\pmb{H}$ are non-positive, the NQP is submodular.

For our experiment, we randomly generate $n\!=\!100$ monotone DR-submodular NQP functions, 
where each $\pmb{H} \in \R^{n\times n}$ is a symmetric matrix sampled uniformly from $[-100, 0]$. We further generated a set of $m \!=\! 50$ linear constraints to construct the positive polytope $\mathbb{P} = \{\x \in \R^{n}, \pmb{Ax} \leq \pmb{b}, 0 \leq \x \leq \bar{\pmb{u}}\} $, where entries in $\pmb{A} \in \R^{m\times n}$ are uniformly sampled from $[0, 1]$, $\bar{\pmb{u}} = \boldsymbol{1}$, and $\pmb{b} = \boldsymbol{1}$. 
To make $f$ monotone, we ensure the gradient of $f$ is non-negative by setting $\pmb{h} = -\pmb{H}\bar{\pmb{u}}$. \looseness=-1

\vspace{-2mm}
\paragraph{Optimal budget allocation with continuous assignments.}
The budget allocation problem can be modeled as a bipartite graph $(S, T; W)$, where $S$ is a set of advertising channels and $T$ is a set of customers. The edge weight $p_{st} \in W$ represent the influence probability of channel $s$ on customer $t$. The objective is to maximize the total influence on the customers by allocating the budget to the set of advertising channels. The total influence on customer $t$ from all the channels can be model by a monotone DR-submodular function $I_t(\x) = 1 - \prod_{(s, t) \in W} (1-p_{st})^{\x_s}$, where $\x_s \in \mathbb{R}_+ $ is the budget allocated to channel $s$. Then for a set of $k$ advertisers, where $\x^i \in \mathbb{R}_+^S$ is the budget allocation of the $i^{th}$ advertiser and $\x = [\x^1, \cdots, \x^k]$, the overall objective is \vspace{-6mm}
\begin{align}
g(\x) = \sum_{i=1}^k \alpha_i f(\x^i) , \quad &\text{with} \quad f(\x^i) = \sum_{t \in T} I_t(\x^i), \\
&0 \leq \x^i \leq \bar{\pmb{u}}^i, \quad
\forall 1 \leq i \leq k, \nonumber
\end{align}
where $\alpha_i$ is a constant weight coefficient and $\bar{u}^i$ is the budget limit on each channel for the $i^{th}$ advertiser.

For a real-world instance of the budget allocation problem, we use the Yahoo! Search Marketing Advertiser Bidding Data \cite{yahoo}, which includes search keyword phrases and the bids placed on them by online customers. The dataset consists of 1,000 search keywords, 10,475 customers and 52,567 edges, where each edge between a keyword and customer represents the customer has bid on the keyword. A customer may bid on one phrase multiple times, and we use the frequency of a (phrase, customer) pair to measure the influence probability of the phrase on that customer. Additionally, we use the average bidding price across all the bids in the dataset as the limit on the budget of all the advertisers. 

\subsection{Bad Solutions and High-probability Bounds}
\begin{figure}[t]
\vspace{-2mm}
    \centering
    \includegraphics[width=0.23\textwidth]{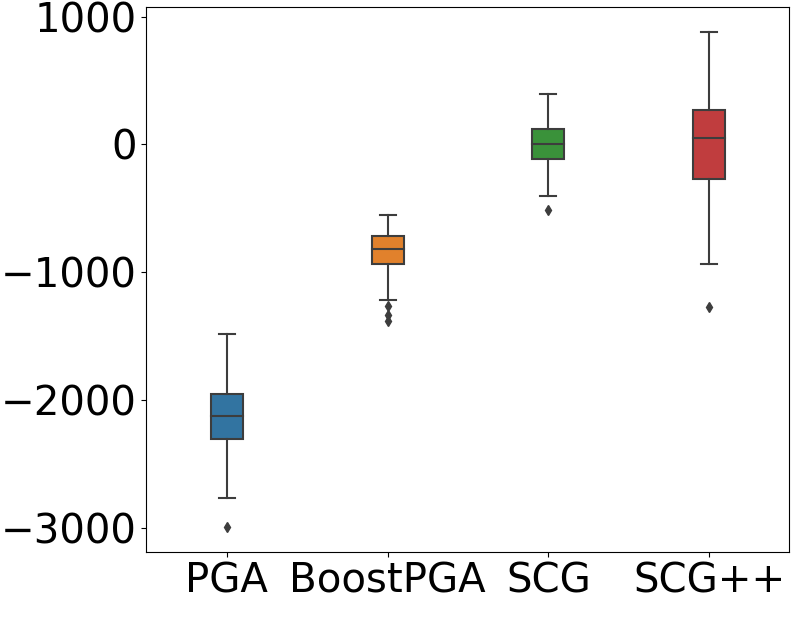}
    \includegraphics[width=0.23\textwidth]{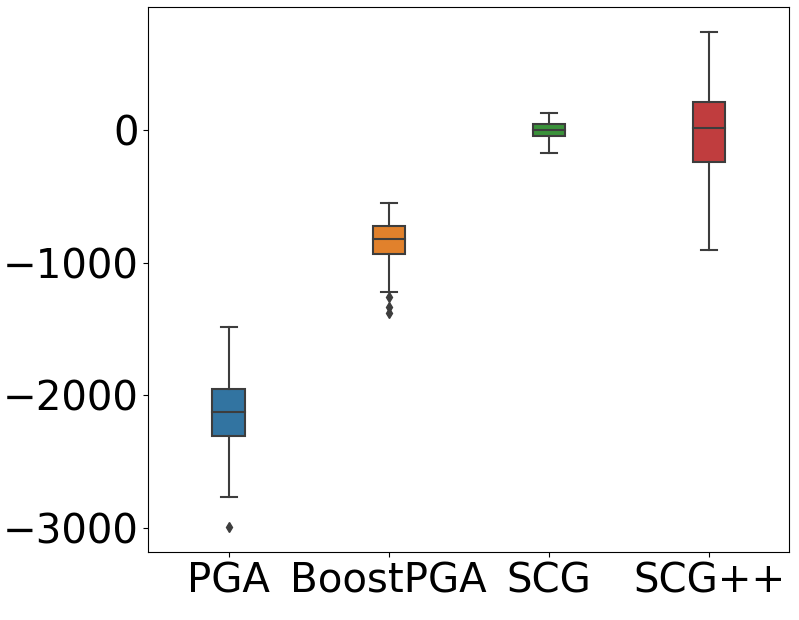}
    \vspace{-1mm}
    \caption{
    The distribution of $\min_{\tau\in [T]} F(\x_\tau) - \mathbb{E}[F(\x_\tau)]$ for PGA and boosted PGA, and $F(\x_T) - \mathbb{E}[F(\x_T)]$ for SCG and SCG++ on NQP, over 100 runs with $T=5$ (left), and $T=100$ (right). 
    Bad solutions 
    can be much worse than the expected values. 
    }\vspace{-2mm}
    \label{fig:notexpect}
\end{figure}
\begin{figure*}[t]
\centering
\begin{subfigure}[NQP, PGA \label{subfig:nqp_pga}]{
\includegraphics[width=.28\textwidth,height=.2\textwidth]{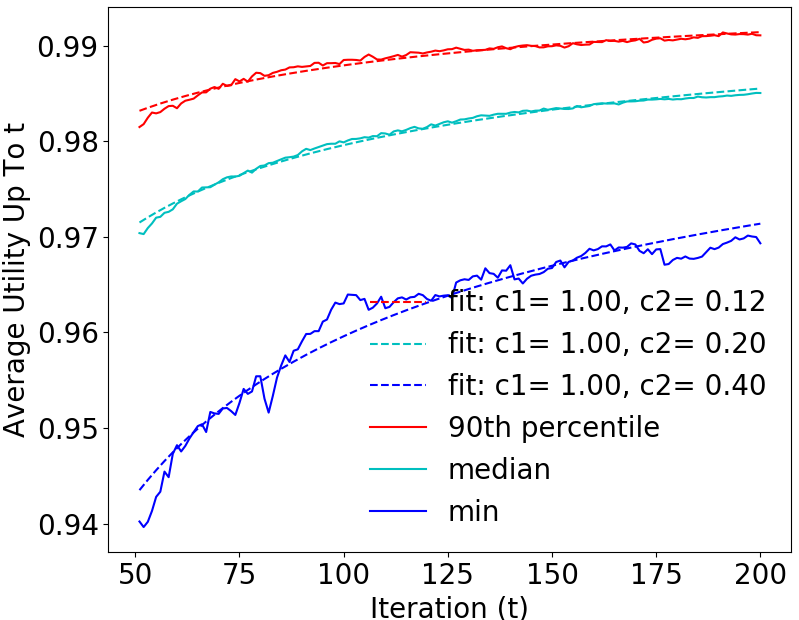}
\vspace{-5mm}
}
\end{subfigure}
\begin{subfigure}[NQP, Boosted PGA \label{subfig:yahoo_boost_pga}]{
\includegraphics[width=.28\textwidth,height=.2\textwidth]{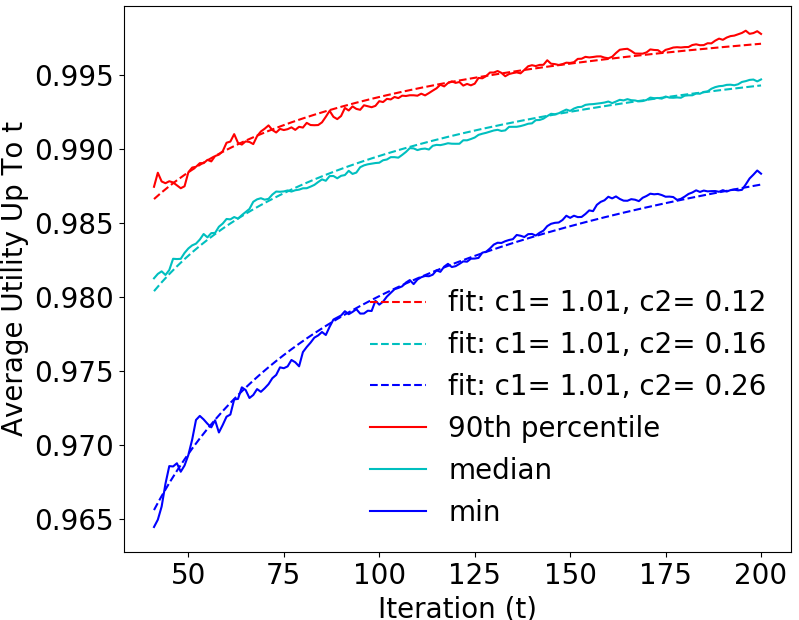}
\vspace{-3mm}
}
\end{subfigure}
\begin{subfigure}[NQP, SCG \label{subfig:nqp_scg}]{
\includegraphics[width=.28\textwidth,height=.2\textwidth]{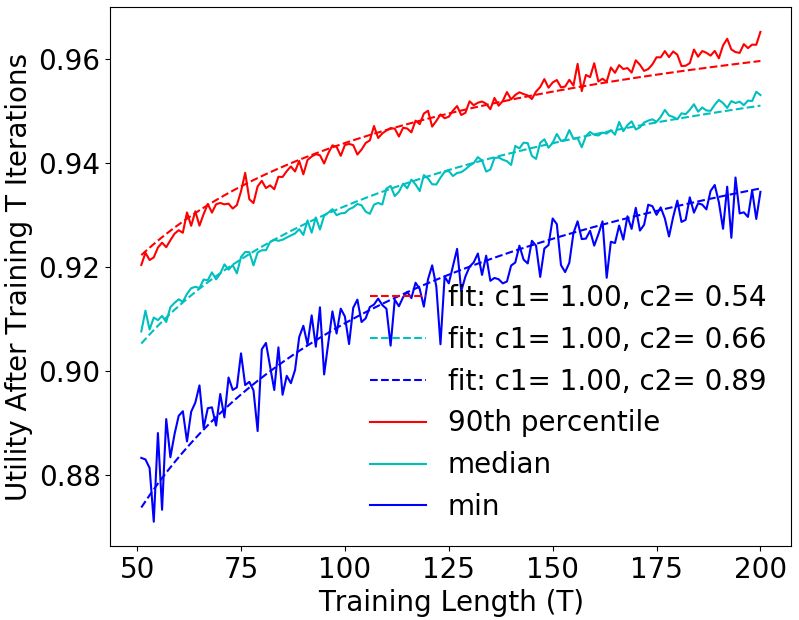}
\vspace{-5mm}
}
\end{subfigure}
\begin{subfigure}[NQP, SCG++ \label{subfig:nqp_scgplus}]{
\includegraphics[width=.28\textwidth,height=.2\textwidth]{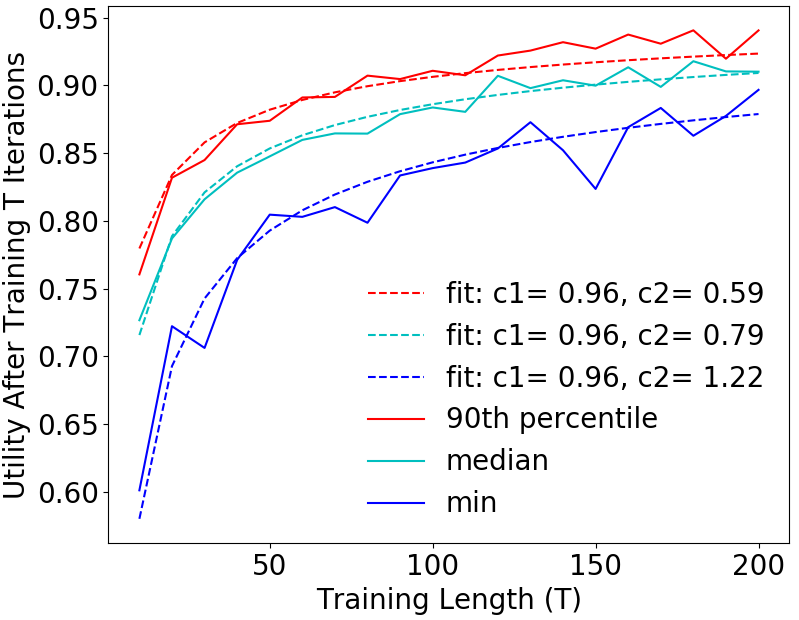}
}
\end{subfigure}
\begin{subfigure}[Yahoo!, PGA \label{subfig:yahoo_pga}]{
\includegraphics[width=.28\textwidth,height=.2\textwidth]{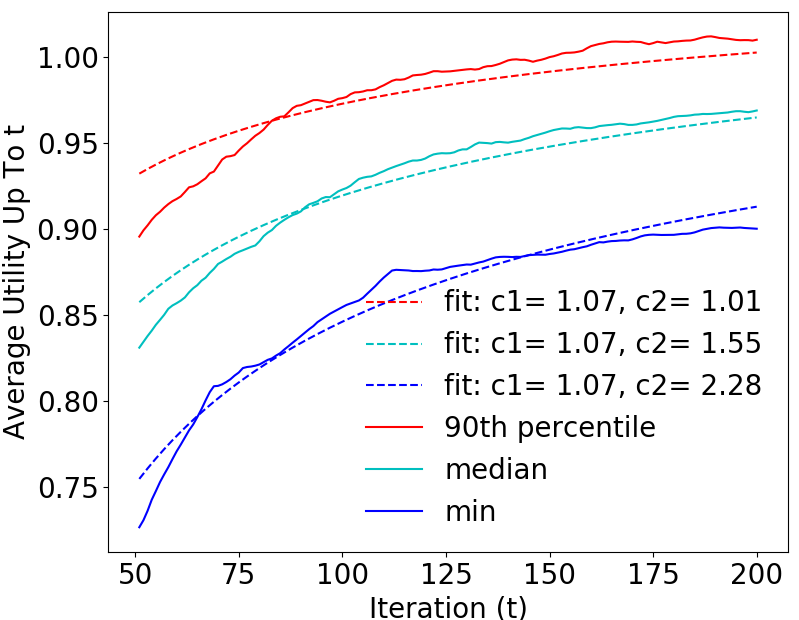}
\vspace{-3mm}
}
\end{subfigure}
\begin{subfigure}[Yahoo!, Boosted PGA \label{subfig:yahoo_boost_pga}]{
\includegraphics[width=.275\textwidth,height=.2\textwidth]{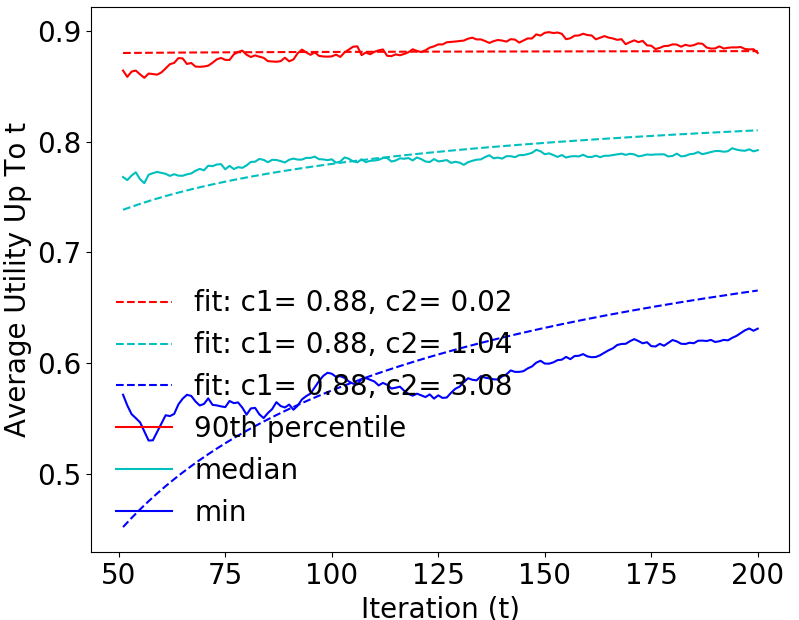}
\vspace{-3mm}
}
\end{subfigure}
\begin{subfigure}[Yahoo!, SCG \label{subfig:yahoo_scg}]{
\hspace{0mm}
\includegraphics[width=.28\textwidth,height=.2\textwidth]{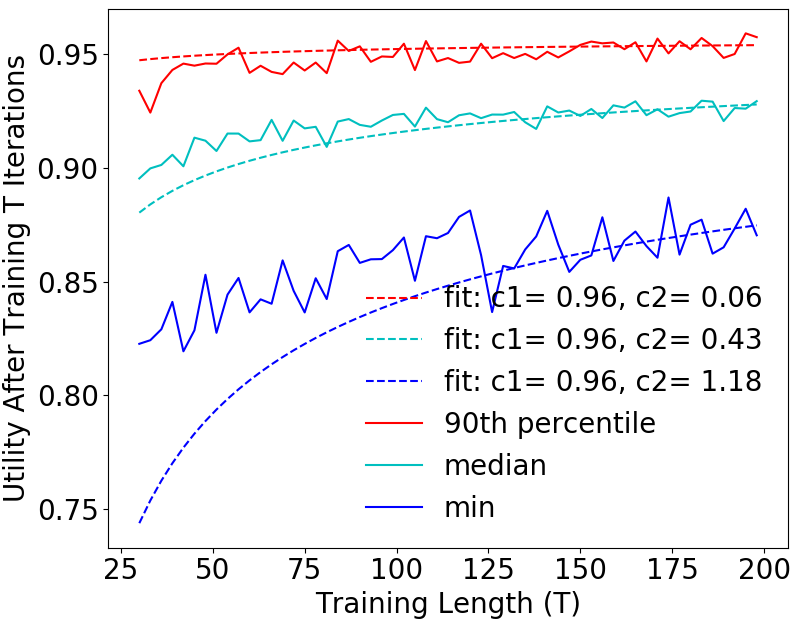}
\vspace{-3mm}
}
\end{subfigure}
\begin{subfigure}[Yahoo!, SCG++ \label{subfig:yahoo_scgplus}]{
\includegraphics[width=.28\textwidth,height=.2\textwidth]{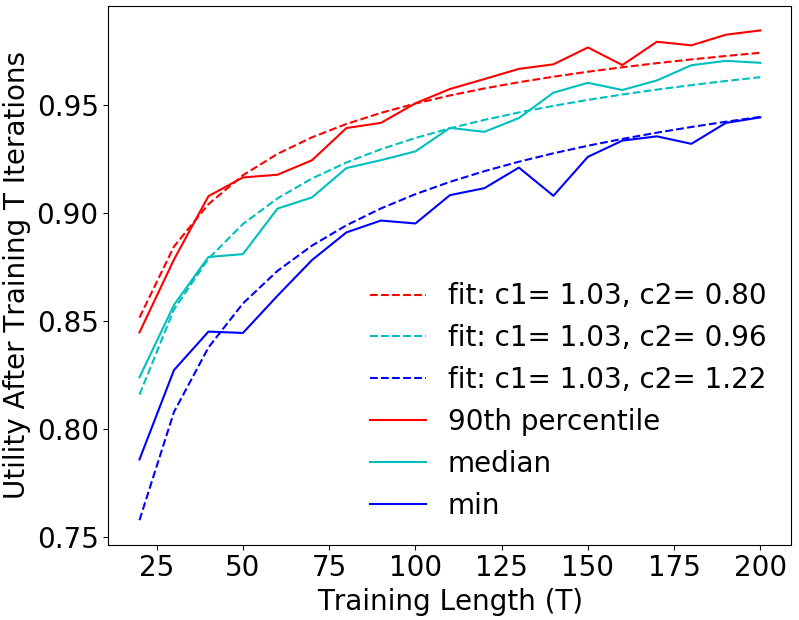}
\vspace{-1mm}
}
\end{subfigure}
\hspace{-2mm}
\begin{subfigure}[NQP, SCG \label{subfig:example}]{
\hspace{-2mm}
\includegraphics[width=.3\textwidth,height=.2\textwidth]{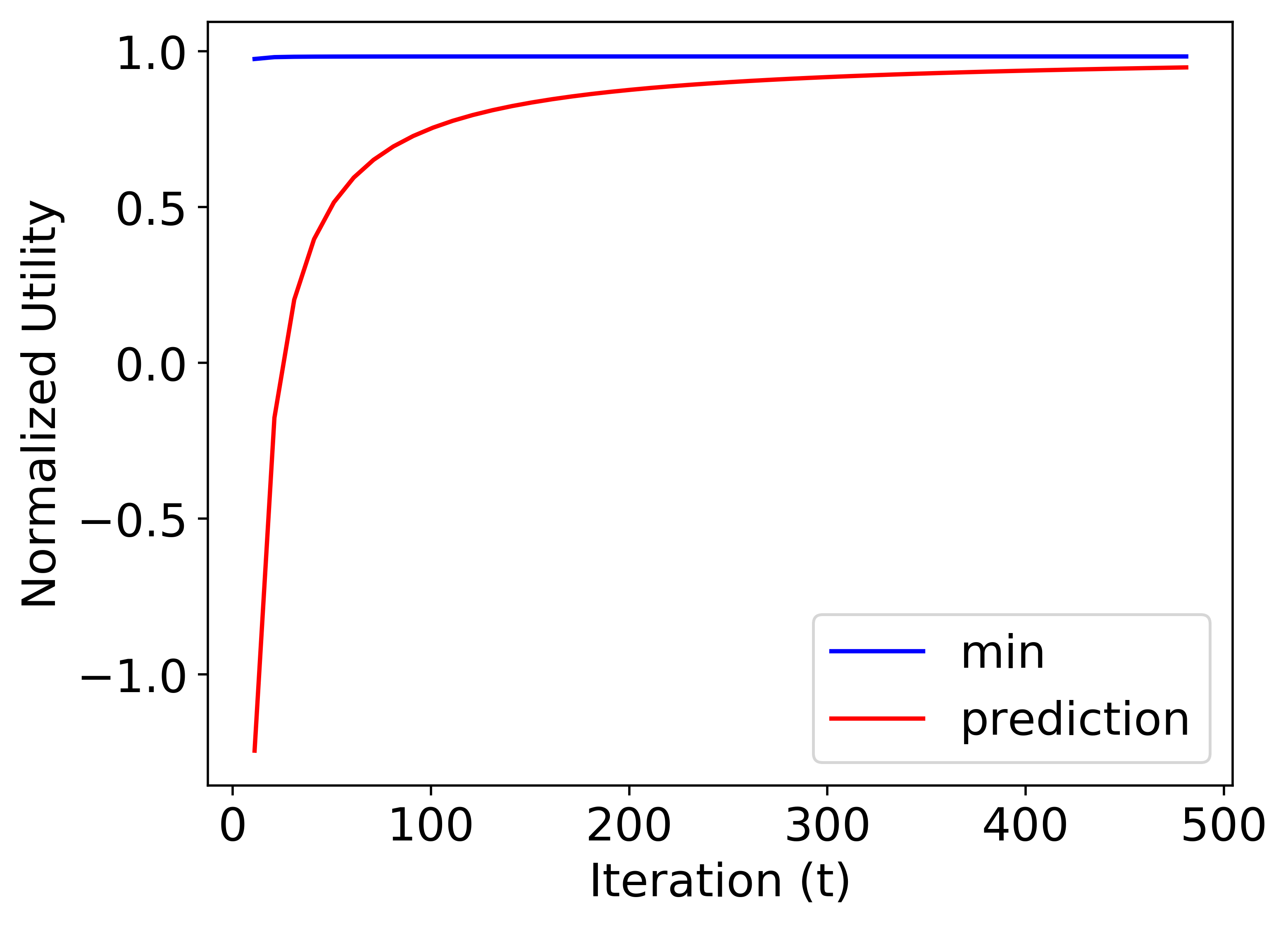}
\vspace{-3mm}
}
\end{subfigure}
\vspace{-2mm}
\caption{(a)-(h) Median, minimum, {and 90\% percentile} of the normalized solutions obtained by PGA, Boosted PGA, SCG, SCG++ compared to our bounds provided in Theorems \ref{thm:pga_bound},
\ref{thm:nonobv_pga_bound}, \ref{thm:scg_strong}, \ref{thm:scg_plus}. The results of PGA, Boosted PGA, SCG, SCG++ are averaged over 100 runs. (i) validating our bounds on a simple NQP example where the constants are known. 
}\vspace{-2mm}
\label{fig:min_med}
\end{figure*}

\vspace{-1mm}
\textbf{Setup.}
We repeat every experiment 100 times, and use a step-size of $1e-4$, 
$1e-2$ for PGA methods on NQP
and Yahoo! respectively. We set the step-size to $1/T$ for SCG and SCG++, 
and use batch size of $T$ for SCG++.
For SCG, we use $\frac{4}{(t+8)^{2/3}}$ as the momentum coefficient. 
For PGA methods, we randomly initialize $\x_0\sim\mathcal{N}(0, 1)$ from a Gaussian distribution, 
and for SCG and SCG++ we initialize $\x_0=\pmb{0}$. Additionally, for PGA, boosted PGA, and SCG experiments, we add a noise sampled from a Gaussian distribution with mean 0 and standard deviation proportional to the gradient norms normalized by its dimensionality to the queried gradients. SCG++ uses noisy estimates of the Hessian, hence we add a smaller Gaussian noise with mean 0 to the Hessian.

\textbf{Bad solutions are far from expectation.}
First, we look at the solution of 
PGA, SCG, and SCG++ on NQP, to see how far the solution may be from the expected value.
Note that when running each algorithm for $T$ iterates, PGA returns the solution for a random iterate $\tau\in[T]$, and SCG, SCG++ return the solution of the final iterate $T$.
Fig. \ref{fig:notexpect} shows the distribution of $\min_{\tau\in [T]} F(\x_\tau)$ for PGA, and $F(\x_T)$ for {boosted PGA,} SCG and SCG++ for $T=5$ (left), and $T=100$ (right), over 100 runs. We see that bad solutions of the algorithms can be much worse than their expected value. 
While more number of iterations reduces the variance of SCG, we see that PGA and SCG++ have a very high variance 
even after $T=100$ rounds (see \Cref{apx:experiments} for more details). 
This shows the insufficiency of the expected guarantees, and confirms the necessity of high-probability analysis. \looseness=-1

\textbf{High-probability bounds.}
Next, we empirically confirm the validity of our high-probability bounds. 
To do so, we first apply PGA, Boosted PGA, SCG, and SCG++ to the continuous submodular problems discussed in Sec. \ref{sec:csp}.
Then, we compare the empirical results with our bounds in Theorems \ref{thm:pga_bound}, \ref{thm:nonobv_pga_bound}, \ref{thm:scg_strong}, \ref{thm:scg_plus}.
Specifically, for each iteration $t$, we report the average utility up to $t$, i.e. $\frac{1}{t}\sum_{i=1}^t F(\x_i)$, for PGA; and the value of $F(\x_t)$ 
for Boosted PGA, SCG, and SCG++.
To avoid the need for calculating the exact scaling constants in Theorems (e.g. $L, D, K$, etc.) and the true optimal value $OPT$ (which determines the asymptote), we fit a line in the same form of the lower bound derived in Theorems \ref{thm:pga_bound}, 
\ref{thm:nonobv_pga_bound}, \ref{thm:scg_strong}, \ref{thm:scg_plus}, to the output of the algorithms. For each line, $c_1$ 
corresponds to fraction of $OPT$ the bound is converging to, while $c_2$ scales the rate of convergence 
depending on the problem-specific constants and desired confidence threshold. 
Specifically, for each algorithm we fit a line of the form $l(t)=c_1\!-\!\frac{c_2}{\sqrt{t}}$.
Importantly, as SCG++ uses batch size 
of $\mathcal{O}(T)$, an equivalent form of this fitted line is $l
(t)=c_1\!-\!\frac{c_2}{k^{1/4}}$, which is slower than the previous two algorithms. 
For $c_1$, we use the same value for min, median, and different percentiles, by taking the average of the $c_1$ values obtained from the fitted lines for an algorithm on the same dataset.
Using the above 
$c_1$, we fit the curves again to get the corresponding $c_2$ for each line.\looseness=-1 

Fig. \ref{fig:min_med}(a)-(h) show the median, minimum, {and 90\% percentile} of utility over the course of training of each algorithm, compared to our predicted lower bounds.
We see that our bounds closely match these utility statistics for various iterations of PGA, SCG, SCG++, and boosted PGA applied to different problems. Since for each percentile level the bound is of the same order, the \emph{differences} in percentiles decrease as well (e.g. $\frac{a}{t^{1/2}} - \frac{b}{t^{1/2}} =  \frac{c}{t^{1/2}}$). This effect can be seen as the minimum returned value across runs approaches the median and 90\% returned values as the number of training iterations increase.

We further validate our bounds by adding a simple example where the constants are known and running SCG on the problem. 
We construct a small NQP where $\pmb{H} \!\in\! \R^{5\times 5}$ is sampled uniformly from $[-1, 0]$. 
Hence, the Lipschitz constant 
$L=\norm{\pmb{H}}_2 \!=\! \sqrt{\lambda_{max}(\pmb{H}^T \pmb{H})}$.
We use linear constraints as Sec. \ref{sec:csp} and set $\pmb{A} = [0.2, 0.2, 0.2, 0.2, 0.2]$.
Thus, diameter $D=\norm{\textbf{1}}_2$. 
For a clipped Gaussian noise $clip(\mathcal{N}(0, \sigma), -2\sigma, 2\sigma)$ to queried gradients for SCG, $M=2\sqrt{5}\sigma$. The optimal value for the problem is approximated by taking the maximum value across 100 SCG runs with 5000 iterations. 
Using above constants in Theorem \ref{thm:pga_bound},
\ref{thm:nonobv_pga_bound}, \ref{thm:scg_strong}, \ref{thm:scg_plus}, Fig. \ref{subfig:example} shows our predicted lower bound with 99\% confidence converges quickly to the minimum of the collected utility trajectories.

%% file: conclusion.tex
\vspace{-1mm}
\section{CONCLUSION}\vspace{-1mm}
We derived the first high probability analysis of the existing methods for stochastic Continuous Submodular Function (CSF) maximization, namely PGA, {boosted PGA}, SCG, and SCG++.
When assumptions on the stochasticity of gradients are strong enough, we showed that even in the worst case the solutions of the algorithms are lower bounded by a function converging to their expected guarantees.
Specifically, with $K$ stochastic gradient computations, we
demonstrated that PGA converges with rate of $O(\frac{1}{\sqrt{K}})$ to the expected $OPT/2$ {and the boosted version at the same rate to $(1-1/e)OPT$.} 
For SCG and SCG++, we showed 
that both algorithms converge at rates of at least $O(\frac{\delta}{K^{\frac{1}{3}}})$ and $O(\frac{\delta}{K^{\frac{1}{2}}})$ to the expected $(1-1/e)OPT$, where $\delta$ depends on the confidence threshold. Besides, under the sub-Gaussian assumption on the gradient noise, we provided an improved lower bound of $O(\frac{1}{\sqrt{K}})$ for the convergence of SCG, that is faster than the existing convergence rate to the expected solution. 
Our results allows characterizing worst
and best-case performance of
CSF maximization in stochastic settings, and 
mitigating the risk of getting a bad solution.

%% file: appendix.tex
\section{ADDITIONAL PROOFS}
\input{proofs/martingale}
\input{proofs/pga}
\input{proofs/pga_non_oblivious}
\input{proofs/scg}

\input{proofs/scg_alt}
\input{proofs/momentum}

\input{proofs/scg_plus}

\section{CONTINUOUS SUBMODULAR MAXIMIZATION ALGORITHMS}\label{apx:alg}

\begin{algorithm}[H]
\SetAlgoLined
 initialization: convex constraint set $\mathcal{C}$ with step lengths $\eta_t$\\
 initialization: $\x_0 \in \mathcal{C}$\\
 \medskip
 \For{t=1....T}{
        $\g_t \gets \nabla F(\x_t, \z_t)$, where $\z_t\sim P$\\
        $\y_{t+1} \gets \x_t + \eta \g_t$\\
        $\x_{t+1} \gets \arg\min_{\x \in C} ||\y_{t+1} - \x ||_2$ \\ 
 }
 \Return $\x_T$
 \caption{Projected Gradient Ascent}
 \label{alg:pga}
\end{algorithm}

\begin{algorithm}[H]
\SetAlgoLined
 initialization: convex constraint set $\mathcal{C}$ with step lengths $\eta_t$\\
 initialization: $\x_0 \in \mathcal{C}$\\
 \medskip
 \For{t=1....T}{
        Sample $s_t$ from $\pmb{S}$ where $\mathbb{P}(\pmb{S}\leq s) = \int_0^s \frac{\gamma^{\gamma(u-1)}}{1-e^{-\gamma}} \mathds{1}(u\in[0,1])du$ \\
        Compute     $\nabla \tilde{F}'(\x_t) := \frac{1-e^{-\gamma}}{\gamma}\nabla \tilde{F}(s_t*\x_t)$\\
        $\y_{t+1} \gets \x_t + \eta \nabla \tilde{F}'(\x_t)$\\
        $\x_{t+1} \gets \arg\min_{\x \in C} ||\y_{t+1} - \x ||_2$ \\ 
 }
 \Return $\x_T$
 \caption{Boosted Projected Gradient Ascent}
 \label{alg:pga}
\end{algorithm}

\begin{algorithm}[H]
\SetAlgoLined
 initialization: Stepsizes $\rho_{t} > 0$.\\ 
 initialization: $\gb_0 = \x_0 = 0$\\
 \medskip
 \For{t=1....T}{
        $\gb_t \gets (1 - \rho_t) \gb_{t-1} + \rho_t \nabla \tilde{F}(x_t,z_t)$ \\
        $\vb_t \gets \arg\max_{\vb \in C}\{\gb_{t}^{T} \vb\}$\\
        $\x_{t+1} \gets \x_{t+1} + \frac{1}{T} \vb_t $ \\ 
 }
 \Return $\x_T$
 \caption{Stochastic Continuous Greedy}
 \label{alg:scg}
\end{algorithm}

\begin{algorithm}[H]\label{alg:scg_plus}
\SetAlgoLined
 input: minibatch sizes $|\mathcal{M}_0|$ and $|\mathcal{M}|$ \\ 
 initialization: $\x^0 = 0$ \\
 \medskip
 \For{t=1....T}{
         \uIf{t=1}{
             $M_0 \sim p(\z;\x^0)$ and find $\gh_0 = \nabla \tilde{F}(x_0, \mathcal{M}_0)$
        }
        \uElse{ 
             $a \sim Unif[0,1]$\\
             $\x(a) = a \cdot \x_t + (1-a) \cdot \x_{t-1}$\\            
             $\mathcal{M} \sim p(\z; \x(a))$\\
             $\widetilde{\nabla_{t}^{2}} = \frac{1}{|\mathcal{M}|} \sum_{(a,\z(a)) \in \mathcal{M}} \widetilde{\nabla^{2}} F(\x(a); \z(a)) $\\
            $\tilde{\Delta^t} = \widetilde{\nabla_{t}^{2}} (\x^t - \x^{t-1})$\\ 
             $\gh_t \gets \gh_{t-1} + \tilde{\Delta^t}$
        } 
        $\vb_t \gets \arg\max_{\vb \in C}\{\gh_{t}^{T} \vb\}$\\
        $\x_{t+1} \gets \x_{t+1} + \frac{1}{T} \vb_t $
 }
 \Return $\x_T$
 \caption{Stochastic Continuous Greedy++}
\end{algorithm}

\section{ADDITIONAL EXPERIMENTAL RESULTS}\label{apx:experiments}
\subsection{Distance Between Expectation and Bad Solutions}
\begin{figure}[H]\label{fig:pga-hist1}
    \centering
    \includegraphics[width=0.35\linewidth]{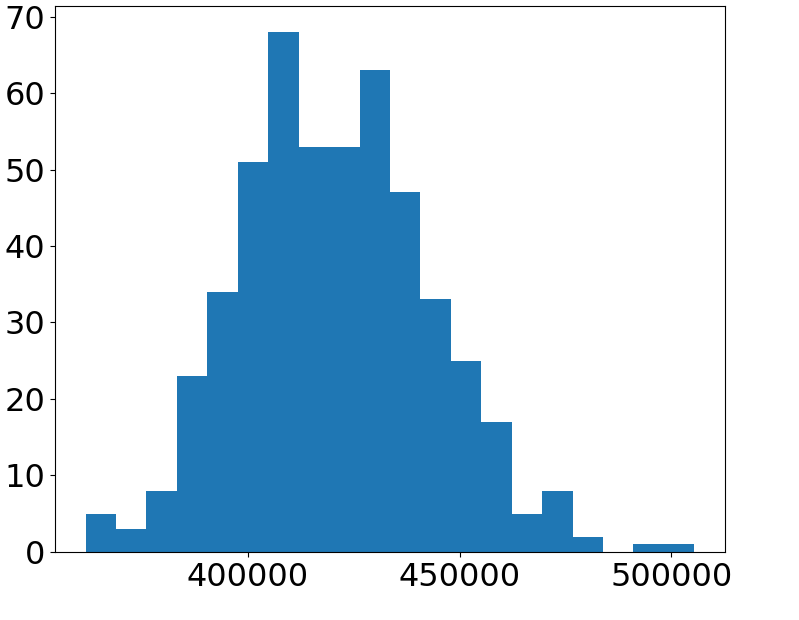}
    \includegraphics[width=0.35\linewidth]{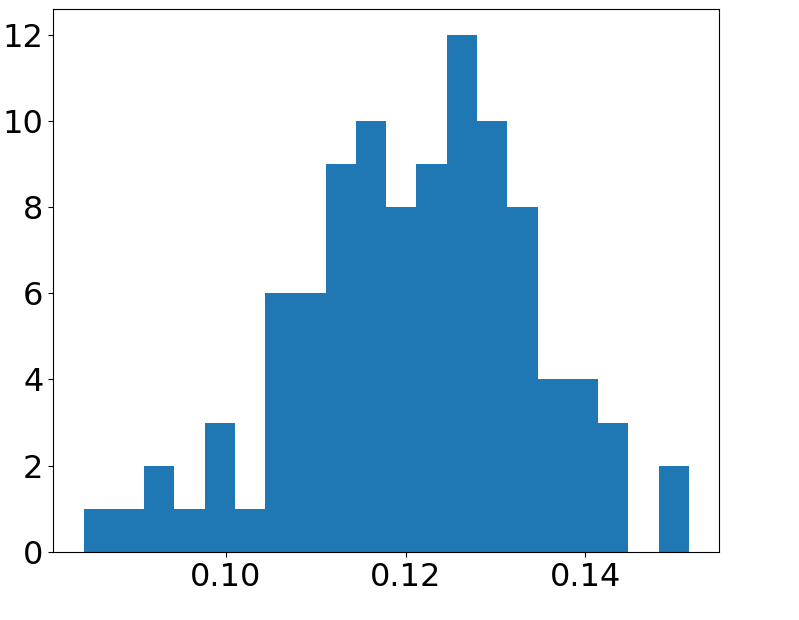}
    \caption{Variance of PGA Utility Distribution. NQP (Left) Yahoo! (Right).}\label{figure:PGA_variance}
\end{figure}

\begin{figure}[H]\label{fig:pga-hist2}
    \centering
    \includegraphics[width=0.35\linewidth]{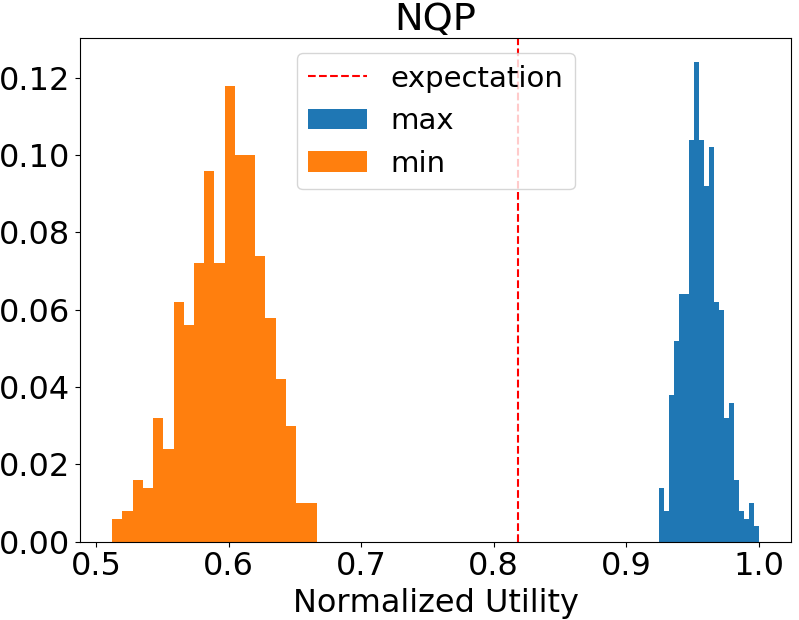}
    \includegraphics[width=0.35\linewidth]{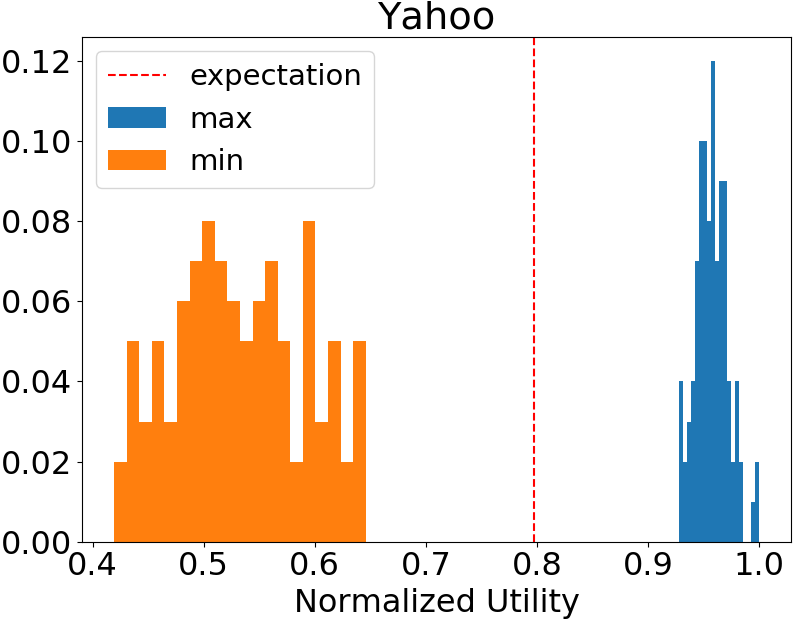}
    \caption{Normalized Utility Distribution of PGA. Histograms of normalized minimum and maximum utility of each individual run, and the expected utility, across 500 runs for NQP \& Yahoo!.}\label{figure:PGA_min_max}
\end{figure}

In this section, we present more results on how far away the worst-case PGA solutions can be from the expectation. In \Cref{figure:PGA_variance} we show the distribution of variance in returned solution utility for each run of the algorithm. \Cref{figure:PGA_min_max} shows the distribution of minimum and maximum \emph{normalized} utilities and the expectations for PGA on each dataset. We observe that there is a large gap between the minimum solutions and the expectation, and in the extreme case a bad run can have 30\% less value than the expectation. This observation demonstrates the need for providing a lower bound on the performance of the existing stochastic continuous submodular maximization methods.

%% file: proofs/martingale.tex
\subsection{Background on Martingale Sequences}
We first provide a brief review on Martingale sequences, which can be found in any standard text on stochastic processes such as \cite{chung2006concentration}. Given probability space $\Omega$ and probability distribution p, we first denote $\mathcal{F}$ as a $\sigma$-field on $\Omega$. We also denote the filtration $\mathbb{F}$ as a nested sequence of $\sigma$-subfields:
\begin{equation}
    \mathbb{F} := \{\mathcal{F}_t\}_{t\leq n} \quad s.t. \quad
    \mathcal{F}_0\subset\mathcal{F}_1\subset\dots \subset \mathcal{F}_n = \mathcal{F}
\end{equation}
We say that a sequence of random variables $\{X_t\}_{t=1}^n:=\{X_1, \dots, X_n\}$ is Martingale with respect to filtration $\mathbb{F}$ if the following properties hold:
\begin{align}
    \mathbb{E}[|X_t|]\leq \infty \nonumber\\
    \mathbb{E}[X_{t+1}|\mathcal{F}_t] = X_t
\end{align}
In other words, given past observations, the next random variable in the sequence is expected to take on the value of the previous one. 

A Martingale \emph{difference} sequence $\{Y_t\}_{t\leq n}$ with respect to filtration $\mathbb{F}$ is defined as having a conditional expectation of zero
\begin{equation}
    \mathbb{E}[Y_{t+1}|\mathcal{F}_t] = 0
\end{equation}
It is easy to see that given a Martingale sequence $\{Y_t\}_{t=1}^n$, one can construct a difference sequence by setting $Y_t=X_t-X_{t-1}$.

Lastly, we say a Martingale is $c$-lipschitz if $\forall t$
\begin{equation}
    |X_t - X_{t-1}| \leq c_t
\end{equation}
This is equivalent to saying all random variables $Y_t$ in the corresponding Martingale difference sequence are bounded.

%% file: proofs/pga.tex
\subsection{Proof of \Cref{thm:pga_bound}}\label{sec:pga_proof}
We first make use of an alternative form  of  the usual Azuma-Hoeffding inequality from \cite{chung2006concentration}: 
\begin{theorem}{(Chung and Lu 2006, theorem 5.2)}\label{thm:chung-concentration}
If martingale X is c-lipschitz, then
\begin{equation}
    \mathbb{P}\left(|X-\mathbb{E}(X)| \leq \sqrt{2\sum_i c_i^2 \log(1/\delta)}\right) \geq 1-\delta
\end{equation}
\end{theorem}
Next, we prove \cref{thm:pga_bound}.
\begin{proof}
Submodularity guarantees that for any two points $x,y \in \mathcal{X}$:
\begin{equation}
    \nabla F(\x) \geq \nabla F(\y) \: \text{for all} \: \x\leq \y
\end{equation}
Using eq 7.2 from Hassani et al. 2017 in \cite{Hassani2017gradient} we know that for any two points $\x, \y \in \mathcal{X}$ we have the relation 
\begin{align}
    F(\y)-2F(\x) \leq \left< \nabla F(\x), \y-\x\right>
\end{align}
This allows us to proceed with a derivation of convergence similar to \cite{zhang2017empirical} for SGD: First replace our $\y,\x$ in the previous inequality with $\x^*$ (the maximizing input to our function) and $\x_t$ (the $t$-th iteration of our algorithm) respectively. Then letting $\hat{\z}_t:= \g_t - \nabla F(\x_t)$ denote the random difference between the true and noisy gradient we have 
\begin{align}\label{eq:proof1-step1}
     F(\x^*)-2 F(\x_t) &\leq  \left< \nabla F(\x_t), \x^*-\x_t\right> \nonumber \\
     &= \left<\x^*-\x_t, \g_t\right> - \left<\x^*-\x_t, \hat{\z}_t\right> \nonumber \\
     &\leq  \frac{1}{\eta_t}\left<\x^*-\x_t, \x'_{t+1}-\x_t\right> - \left<\x^*-\x_t, \hat{\z}_t\right>
\end{align}
In the second inequality we use the definition of our gradient step (before projection) $\x'_{t+1} = \x_t + \eta_t \g_t$. Next, through some algebra we get
\begin{align}\label{eq:proof1-step2}
     F(\x^*)-2F(\x_t) &\leq \frac{1}{2\eta_t}(\norm{\x_t-\x'_{t+1}}_2^2 + \norm{\x_t-\x^*}_2^2 - \norm{\x'_{t+1}-\x^*}_2^2)- \left<\x^*-\x_t,\hat{\z}_t\right> \nonumber \\
     &\leq \frac{1}{2\eta_t }(\norm{\x_t-\x'_{t+1}}_2^2 + \norm{\x_t-\x^*}_2^2 - \norm{\x_{t+1}-\x^*}_2^2)- \left<\x^*-\x_t, \hat{\z}_t\right>
\end{align}
By the property of euclidean projections we know $\norm{\x_{t+1}-\x^*} \leq \norm{\x'_{t+1}-\x^*}$ which gives us the second line above. Next we use our gradient step equation and lipschitz-smoothness assumption to get:
\begin{align}\label{eq:proof1-step3}
    F(\x^*)-2F(\x_t) &\leq \frac{\eta_t}{2}\norm{\g_t}_2^2 + \frac{1}{2\eta_t}(\norm{\x_t-\x^*}_2^2 - \norm{\x_{t+1}-\x^*}_2^2)- \left<\x^*-\x_t, \hat{\z}_t\right> \nonumber \\
    &\leq \frac{\eta_t (L+M)^2}{2} + \frac{1}{2\eta_t }(\norm{\x_t-\x^*}_2^2 - \norm{\x_{t+1}-\x^*}_2^2)- \left<\x^*-\x_t, \hat{\z}_t\right>
\end{align}
We let $\Delta_t \triangleq \left<\x_t-\x^*, \hat{\z}_t \right>$, and $\eta_t = \frac{2}{\sqrt{t}}$ to get
\begin{align}\label{eq:single-point-bound}
    F(\x^*)-2F(\x_t) &\leq \frac{(L+M)^2}{\sqrt{t}} + \frac{\sqrt{t}}{4}(\norm{\x_t-\x^*}_2^2 - \norm{\x_{t+1}-\x^*}_2^2) + \Delta_t \nonumber \\
     &\leq \frac{(L+M)^2}{\sqrt{t}} + \frac{\sqrt{T}}{4}(\norm{\x_t-\x^*}_2^2 - \norm{\x_{t+1}-\x^*}_2^2) + \Delta_t
\end{align}
If combine the inequalities for each $\x_t$ and divide by T we have:
\begin{equation}
     F(\x^*) - (\frac{2}{T})\sum_{t=1}^T F(\x_t) \leq \frac{1}{T} \left( \sum_{t=1}^T\frac{1}{\sqrt{t}}(L+M)^2 + \frac{\sqrt{T}}{4}\norm{\x^*-\x_1}_2^2 + \sum_{t=1}^T \Delta_t\right) \nonumber \\
\end{equation}
Using the fact that $\sum_{t=1}^T \frac{1}{\sqrt{t}} \leq 2\sqrt{T}$ we simplify to
\begin{align}
    F(\x^*) - (\frac{2}{T})\sum_{t=1}^T F(\x_t) &\leq \frac{1}{T} \left( 2\sqrt{T}(L+M)^2 + \frac{\sqrt{T}}{4}\norm{\x^*-\x_1}_2^2 + \sum_{t=1}^T \Delta_t\right) \nonumber \\ 
    &= \frac{2(L+M)^2}{\sqrt{T}} + \frac{\norm{\x^*-\x_1}_2^2}{4\sqrt{T}}  + \frac{1}{T} \sum_{t=1}^T \Delta_t
\end{align}
Rearranging this inequality we see
\begin{align}\label{eq:proof1-combine-step}
    \frac{1}{2}OPT - \frac{1}{T}\sum_{t=1}^T F(\x_t) \leq  \frac{(L+M)^2}{\sqrt{T}} + \frac{\norm{\x^*-\x_1}_2^2}{8\sqrt{T}} + \frac{1}{2T} \sum_{t=1}^T \Delta_t
\end{align}
Let $\Delta'_t \triangleq \frac{1}{2}\Delta_t$ so we can simplify to 
\begin{align}
    \frac{1}{2}OPT - \frac{1}{T}\sum_{t=1}^T F(\x_t) \leq  \frac{1}{\sqrt{T}}\left(\frac{8(L+M)^2 +\norm{\x^*-\x_1}_2^2}{8}\right) + \frac{1}{T} \sum_{t=1}^T \Delta'_t
\end{align}
Our next step is to use the Azuma-Hoeffding inequality  to bound $\frac{1}{T} \sum_{t=1}^T \Delta'_t$. If we can show that $\{\Delta'_t\}_T$ is a bounded martingale difference sequence with zero expectation, then we know by \Cref{thm:chung-concentration} with probability $1-\delta$:
\begin{align}
    |\sum_{t=1}^T \Delta'_t| \lesssim \sqrt{T\log(1/\delta)}
\end{align}
Expanding $\mathbb{E}(\Delta'_t)$ we see that because $\x_t-\x^*$ is independent of the gradient error $\hat{\z}_t$ we have
\begin{align}
    \mathbb{E}(\Delta'_t) &= \mathbb{E}\left(\frac{1}{2}\left<\x_t-\x^*, \hat{\z}_t \right>\right) = \frac{1}{2}\left<\mathbb{E}(\x_t-\x^*), \mathbb{E}(\hat{\z}_t) \right> = 0
\end{align}
Similarly, we can use triangle inequalities and our assumptions to bound $\norm{\Delta'_t}$ by a constant
\begin{align}
    \norm{\Delta'_t} \leq \frac{1}{2}\norm{\x_t-\x^*} \norm{\hat{\z}_t} \leq \frac{DM}{2}
\end{align}
Recall D is the maximum distance between any two points in our set $\mathcal{C}$. Therefore we have
\begin{equation}
    \mathbb{P}\left(|X-\mathbb{E}(X)| \leq DM\sqrt{T \log(1/\delta)/2}\right) \geq 1-\delta
\end{equation}

Therefore we have a martingale difference sequence which implies with probability $1-\delta$
\begin{align}
     \frac{1}{2}OPT - \frac{1}{T}\sum_{t=1}^T F(\x_t) &\leq \frac{1}{\sqrt{T}}\left(\frac{8(L+M)^2 +D^2}{8}\right) + \sqrt{\frac{ \log(\frac{1}{\delta})}{2T}}DM \nonumber \\
     &= O(\frac{1}{\sqrt{T}})
\end{align}
\end{proof}

%% file: proofs/pga_non_oblivious.tex
\subsection{Proof of \Cref{thm:nonobv_pga_bound}}\label{apx:nonobv-proof}
In this section, we demonstrate how to adapt the PGA bound in \cref{thm:pga_bound} to the Boosted PGA setting. For the proof we use the slightly more general form of the non-oblivious function in \cite{zhang2022stochastic} which also works for weakly submodular functions with parameter $\gamma$:
\begin{align}
    F'(\x) := \int_0^1 \frac{e^{\gamma(s-1)}}{s}F(s*\x) ds \\
    \nabla F'(\x) := \int_0^1 e^{\gamma(s-1)}F(s*\x) ds
\end{align}
The noisy gradient estimate of this auxiliary function is now:
\begin{align}\label{eq:nonobv-noisy}
    \nabla \tilde{F}'(\x_t) := \frac{1-e^{-\gamma}}{\gamma}\nabla \tilde{F}(s_t*\x_t).
\end{align}
Here $s_t$ is independently sampled from a distribution with cdf: $\mathbb{P}(\pmb{S}\leq s) = \int_0^s \frac{\gamma^{\gamma(u-1)}}{1-e^{-\gamma}} \mathds{1}(u\in[0,1])du$, where $\mathds{1}$ is the indicator function.\\
We now begin with a slightly stricter version of proposition 1 from \cite{zhang2022stochastic}.

\begin{lemma}[Proposition 1, Zhang et al. 2022]\label{prop:nonobv-martingale}
If $z_t$ is sampled according to the distribution in Eq. \eqref{eq:nonobv-noisy}, then given Assumptions \ref{ass:x_bound},\ref{ass:g_lip},\ref{ass:g_bound} are satisfied (i.e. Lipschitz-smooth function along with bounded domain and gradient noise), we have
\begin{align}
    \mathbb{E}\left(\frac{1-e^{-\gamma}}{\gamma}\nabla \tilde{F}(s_t*\x_t)\right) = \nabla F'(\x) \\
    \norm{\nabla F'(\x) - \nabla \tilde{F}'(\x)}^2 \leq M_{\gamma}^2
\end{align}
where $M_\gamma := (M+2 LD)\left(\frac{1-e^{-\gamma}}{\gamma}\right)$.
\end{lemma}
The first equality statement is unchanged from the original proposition, while the second inequality is modified to bound the squared norm error instead of the variance. We provide a short proof of the second inequality below:
\begin{proof}

\begin{align}
    \norm{\nabla F'(\x) - \nabla \tilde{F}'(\x)} &\leq \norm{\frac{1-e^{-\gamma}}{\gamma}(\nabla \tilde{F}(z*\x)- \nabla F(z*\x))} + \norm{\frac{1-e^{-\gamma}}{\gamma}\nabla F(z*\x) - \nabla F'(x)} \nonumber \\
    & \leq \left(\frac{1-e^{-\gamma}}{\gamma}\right) M + \norm{\frac{1-e^{-\gamma}}{\gamma}\nabla F(z*\x) - \nabla F'(x)}
\end{align}
We use the definition of $\nabla \tilde{F}'(\x)$, as well as the triangle inequality in the first step. In the second step, we use the assumption that the noise of our gradient oracle is bounded by $M$. Finally to bound the second term:
\begin{align}
    \norm{\frac{1-e^{-\gamma}}{\gamma}\nabla F(z*\x) - \nabla F'(x))} &= \norm{\int_0^1 e^{\gamma(u-1)}(\nabla F(z*\x)-\nabla F(u*\x) du} \nonumber \\
    &\leq \left(\int_0^1 e^{\gamma(u-1)}|z-u| L \norm{\x}|\right) \nonumber \\
    &\leq 2LD\left(\int_0^1 e^{\gamma(u-1)}du\right) = 2LD\left(\frac{1-e^{-\gamma}}{\gamma}\right)
\end{align}
We use the Lipschitz smoothness constraint and Cauchy-Schwarz in the first inequality, and $|z-u|<2$ in the second. 
\end{proof}
Now equipped with Lemma \ref{prop:nonobv-martingale}, we can proceed using the same proof structure as in Theorem \ref{thm:pga_bound}:
\begin{proof}
 Starting from $\langle \y-\x, \nabla F'(\x)\rangle \geq (1-e^{-\gamma})F(\y)-F(\x)$ evaluated at $\y=\x^*$, we can show using steps analogous to \cref{eq:proof1-step1,eq:proof1-step2,eq:proof1-step3} the following inequality:
 \begin{align}
     (1-e^{-\gamma})F(\y)-F(\x) &\leq \frac{\eta_t}{2}\norm{\nabla\tilde{F}'(\x_t)}^2 + \frac{1}{2\eta_t} \left(\norm{\x_t-\x^*}^2 - \norm{\x_{t+1}-\x^*}^2 \right) - \langle \x^*-\x_t,\nabla\tilde{F}'(\x_t)- \nabla F'(\x_t)\rangle \nonumber \\
     &\leq \frac{\eta_t}{2}(L_\gamma D + M_\gamma)^2 + \frac{1}{2\eta_t} \left(\norm{\x_t-\x^*}^2 - \norm{\x_{t+1}-\x^*}^2 \right) - \langle \x^*-\x_t,\nabla\tilde{F}'(\x_t)- \nabla F'(\x_t)\rangle 
 \end{align}
 From \cite{zhang2022stochastic} we know that $F'$ is $L_\gamma$-smooth with $L_\gamma = L \frac{\gamma+e^{-\gamma}-1}{\gamma^2}$, giving us the second inequality above. Using $\eta_t = 1/\sqrt{t}$ and applying the bound iteratively produces the parallel equation to \cref{eq:proof1-combine-step}:
 \begin{align}
     (1-e^{-\gamma})OPT-\frac{1}{T}F(\x_t) \leq \frac{1}{\sqrt{T}}(L_\gamma D + M_\gamma)^2 + \frac{\norm{\x^*-\x_1}^2}{8\sqrt{T}} + \frac{1}{2T}\sum_{t=1}^T \langle \x^*-\x_t,\nabla\tilde{F}'(\x_t)- \nabla F'(\x_t)\rangle
 \end{align}
From Lemma \ref{prop:nonobv-martingale} we can treat the summation of $\langle \x^*-\x_t,\nabla\tilde{F}'(\x_t)- \nabla F'(\x_t)\rangle$ terms as a c-lipschitz Martingale with constant:
\begin{equation}
    |\langle \x^*-\x_t,\nabla\tilde{F}'(\x_t)- \nabla F'(\x_t)\rangle| \leq \norm{\x^*-\x_t}\norm{\nabla\tilde{F}'(\x_t)- \nabla F'(\x_t)} \leq DM_{\gamma} =: c
\end{equation}
Therefore, the analogous bound to \cref{thm:pga_bound} for non-oblivious PGA is:
\begin{equation}
    (1-e^{-\gamma})OPT-\frac{1}{T}F(\x_t) \leq \frac{1}{\sqrt{T}} \left(\frac{8(L_\gamma D + M_\gamma)^2+D^2}{8}\right) + \sqrt{\frac{\log(\frac{1}{\delta})}{2T}}DM_\gamma
\end{equation}
 With probability $p>1-\delta$. By substituting $\gamma=1$, we recover the bound for the fully submodular setting.
 
\end{proof}

%% file: proofs/scg.tex
\subsection{Proof of \Cref{thm:scg_bound}}\label{apx:thm2}
We first make use the following lemma from Mokhtari et al. which bounds the variance in momentum error:
\begin{lemma}{(Mokhtari et al. 2018, Lemma 2)}
Given a momentum parameter $p_t = \frac{4}{(t+8)^{2/3}}$, the gradient estimates $\gb_t$ have the following property:
\begin{equation*}
    \mathbb{E}\left[\norm{\nabla F(\x_t) - \gb_t}^2\right] \leq \frac{Q}{(t+9)^{2/3}}
\end{equation*}
\end{lemma}
We note that as a corollary to this lemma, the Chebyshev inequality gives us the following high probability bound:
\begin{equation}
    \mathbb{P}\left(\norm{\nabla F(\x_t) - \gb_t} \geq \delta\sqrt{\frac{Q}{(t+9)^{2/3}}}\right) \leq \frac{1}{\delta^2}
\end{equation}
Squaring both terms inside the condition:
\begin{equation}
    \mathbb{P}\left(\norm{\nabla F(\x_t) - \gb_t}^2 \geq \delta^2\frac{Q}{(t+9)^{2/3}}\right) \leq \frac{1}{\delta^2}
\end{equation}
Now we are ready to begin the proof of \cref{thm:scg_bound}.
\begin{proof}
Starting with the following inequality derived in \cite{mokhtari2018conditional} (Mokhtari et al. 2018, eq. 20)  we have:
\begin{equation*}
    F(\x_{t+1}) - F(\x_t) \geq \frac{1}{T}(F(\x^*)-F(\x_t)) - \frac{LD^2}{2T^2} - \frac{1}{2T}(4\beta_t D^2 + \frac{\norm{\nabla F(\x_t) - \gb_t}^2}{\beta_t})
\end{equation*}

Instead of taking the expectations of both sides as in \cite{mokhtari2018conditional}, we immediately apply the inequality recursively from $t=0,1,\dots,T-1$ to get
\begin{equation*}
    F(\x^*)-F(\x_T) \leq (1-\frac{1}{T})^T (F(\x^*)-F(\x_0)) + \sum_{t=0}^{T-1} \frac{1}{2T}\left[4\beta_tD^2 + \frac{\norm{\nabla F(\x_t) - \gb_t}^2}{\beta_t}\right] + \frac{LD^2}{2T^2} 
\end{equation*}
Looking at the term $\sum_{t=0}^{T-1}\norm{\nabla F(\x_t) - \gb_t}^2$, we can use the union bound as follows:
\begin{equation*}
    \mathbb{P}(\sum_{t=0}^{T-1}\norm{\nabla F(\x_t) - \gb_t}^2 \geq \delta^2 \sum_{t=0}^{T-1}\frac{Q}{(t+9)^{2/3}}) \leq \sum_t \frac{1}{\delta^2} = \frac{T}{\delta^2}
\end{equation*}
Therefore we know that with probability at least $1-\frac{T}{\delta^2}$ that the inequality in the opposite direction is true.

Then choosing $\beta_t = \frac{\delta \sqrt{Q}}{2D(t+9)^{1/3}}$, we get the upper bound:
\begin{equation}
        F(\x^*)-F(\x_T) \leq \frac{1}{e} (F(\x^*)-F(\x_0)) + \sum_{t=0}^{T-1} \delta \frac{2Q^{1/2}D}{(t+9)^{1/3}T} + \frac{LD^2}{2T^2}
\end{equation}
Therefore with probability greater than $1-\frac{T}{\delta^2}$ we have the following bound:
\begin{equation}
        F(\x^*)-F(\x_T) \leq \frac{1}{e} (F(\x^*)-F(\x_0)) +  \delta\frac{2Q^{1/2}D}{T^{1/3}} + \frac{LD^2}{2T^2} 
\end{equation}
Some more manipulation, and we have with probability greater than $1-\frac{T}{\delta^2}$
\begin{equation}
    F(\x_T) \geq (1-\frac{1}{e})F(\x^*) - \delta\frac{2Q^{1/2}D}{T^{1/3}} -\frac{LD^2}{2T^2}
\end{equation}
\end{proof}

%% file: proofs/scg_alt.tex
\subsection{Proof of  \Cref{thm:scg_strong} (Stronger SCG Bound)}\label{sec:scg_strong}
Before verifying \cref{thm:scg_strong}, we first provide the following lemma, which gives a high probability bound for the sum of momentum errors:
\begin{lemma}\label{lem:mom_bound}
Given i.i.d. errors $\norm{\nabla F(\x_t)-\g_t}$ that are each sub-Guassian with parameter $\sigma$, and momentum $\gb_t = (1-\rho_t)\gb_{t-1} + \rho_t \nabla \tilde{F}(\x_t,\z_t)$ with parameter of the form $\rho_t =\frac{1}{t^\alpha}$ and $\alpha \in (0,1)$, then with probability greater than $1-\delta$
 \begin{equation}
     \sum_{t=1}^T\norm{\nabla F(\x_t)-\gb_t} \leq  \sqrt{2K^2\sigma^2T\log(1/\delta)} + LDK
 \end{equation}
 Where $K:=\frac{1}{1-\alpha}\Gamma(\frac{1}{1-\alpha})$
\end{lemma}
{The cumulative error bound in \cref{lem:mom_bound} is to our knowledge the first such result for adaptive momentum optimization methods (i.e. the momentum can change over time). Notably, it is general enough to be used even in the context of other smooth function classes. Recently, \cite{sun2021training} showed adaptive momentum enjoys some superior convergence and generalization properties, and hence our lemma is of independent interest. See \cref{apx:mom_error} for the proof of this lemma. }

Next, we show a tighter high probability bound on the final iterate of the SCG algorithm in \cref{thm:scg_strong}. 
\begin{proof}
Using the smoothness of F and boundedness of $x\in\mathcal{C}$, Hassani et al. 2017 showed that
\begin{equation}
        F(\x_{t+1}) - F(\x_t) \geq \frac{1}{T}(F(\x^*)-F(\x_t)) - \frac{LD^2}{2T^2} + \frac{1}{T}\left<\vb_t-\x^*, \nabla F(\x_t)-\gb_t\right>
\end{equation}
The Cauchy-Schwarz inequality allows us to bound the last term by
\begin{align}
    \frac{1}{T}\left<\vb_t-\x^*, \nabla F(\x_t)-\gb_t\right> \geq -\frac{1}{T}\norm{\vb_t-\x^*}\norm{\nabla F(\x_t)-\gb_t} \geq -\frac{2D}{T}\norm{\nabla F(\x_t)-\gb_t}
\end{align}
Substituting this bound and rearranging terms we have
\begin{equation}
        F(\x^*) - F(\x_{t+1}) \leq (1-\frac{1}{T})(F(\x^*)-F(\x_t))  +\frac{2D}{T}\norm{\nabla F(\x_t)-\gb_t} + \frac{LD^2}{2T^2}
\end{equation}
Applying this inequality recursively for $t=0, \dots, T-1$ we have
\begin{align}\label{eq:pre-lem5}
    F(\x^*) - F(\x_{T}) \leq (1-\frac{1}{T})^T(F(\x^*)-F(\x_0)) + \frac{2D}{T}\sum_{t=0}^{T-1}\norm{\nabla F(\x_t)-\gb_t} + \frac{LD^2}{2T}
\end{align}
Using a momentum term of $\rho_t=\frac{1}{t^{1/2}}$ for \cref{lem:mom_bound} means that $\alpha=0.5$ and hence $K=2$. Directly substituting this inequality into \cref{eq:pre-lem5} gives us
\begin{align}
        F(\x^*) - F(\x_{T}) &\leq (1-\frac{1}{T})^T(F(\x^*)-F(\x_0)) + \frac{2D}{T} \left[ \sqrt{2K^2\sigma^2T\log(1/\delta)} + LDK\right]+ \frac{LD^2}{2T} \nonumber\\
        &\leq \frac{1}{e}(F(\x^*)-F(\x_0)) + \frac{2D\sqrt{2K^2\sigma^2\log(1/\delta)}}{\sqrt{T}} + (\frac{4K+1}{2})\frac{LD^2}{T}
\end{align}
Dropping $F(x_0)$ and rearranging we have
\begin{equation}
    F(\x_T) \geq (1-\frac{1}{e})F(\x^*) - \frac{2DK\sigma\sqrt{\log(1/\delta)}}{T^{1/2}} - (\frac{4K+1}{2})\frac{LD^2}{T}
\end{equation}
\end{proof}

%% file: proofs/momentum.tex
\subsection{Proof of \Cref{lem:mom_bound}}\label{apx:mom_error}
Our first goal is to bound the term $\norm{\nabla F(\x_t) -\gb_t}$. Substituting $\gb_t = (1-\rho_t) \gb_{t-1} + \rho_t \nabla \tilde{F}(\x_t,\z_t)$ we have
\begin{align}
    \norm{\nabla F(\x_t) -\gb_t} = \norm{\nabla F(\x_t) - (1-\rho_t) \gb_{t-1} - \rho_t \nabla \tilde{F}(\x_t,\z_t)}
\end{align}
Adding and subtracting $(1-\rho_t)\nabla F(\x_{t-1})$ on the right hand side  and applying the triangle inequality gives us
\begin{align}
    \norm{\nabla F(\x_t) -\gb_t} = \norm{\rho_t(\nabla F(\x_t) - \nabla\tilde{F}(\x_t,\z_t)) - (1-\rho_t) (\nabla F(\x_t) - \nabla F(\x_{t-1})) + (1-\rho_t) (\nabla F(\x_{t-1}) -\gb_{t-1})} \nonumber \\
    \leq \rho_t \norm{\nabla F(\x_t) - \nabla\tilde{F}(\x_t,\z_t)} + (1-\rho_t)\norm{\nabla F(\x_t) - \nabla F(\x_{t-1})} + (1-\rho_t)\norm{\nabla F(\x_{t-1})-\gb_{t-1}}
\end{align}
Now using our assumptions on the smoothness of F and boundedness of our domain, we know that $\norm{\nabla F(\x_t) - \nabla F(\x_{t-1})} \leq L\norm{\x_t-\x_{t-1}} = L\norm{\frac{\vb_t}{T}} \leq \frac{LD}{T}$ and hence 
\begin{align}\label{eq:mom_recursive}
    \norm{\nabla F(\x_t) -\gb_t} & \leq
    \rho_t \norm{\nabla F(\x_t) - \nabla\tilde{F}(\x_t,\z_t)} + (1-\rho_t)\frac{LD}{T} + (1-\rho_t)\norm{\nabla F(\x_{t-1})-\gb_{t-1}}
\end{align}
Define $a_i := \norm{\nabla F(\x_i) - \nabla\tilde{F}(\x_i,\z_i)}$ as a random variable and recursively apply the inequality in \cref{eq:mom_recursive} to get
\begin{align}
    \norm{\nabla F(\x_t) -\gb_t} \leq  \sum_{i=1}^{t} \rho_i \left(\prod_{j=i+1}^{t}(1-\rho_j) \right) a_{i} + \frac{LD}{T} \sum_{i=1}^{t} \prod_{j=i}^{t}(1-\rho_j) 
\end{align}
Given any sequence of momentum terms $\rho_j$ that is monotonic non-increasing  we have the upper bound $\rho_j \leq \rho_1$ and $(1-\rho_j) \leq (1-\rho_t)$ and therefore
\begin{align}
    \norm{\nabla F(\x_t) -\gb_t} \leq \rho_1 \sum_{i=1}^{t}(1-\rho_t)^{t-i} a_{i} + \sum_{i=1}^{t}(1-\rho_t)^{t-i+1} \frac{LD}{T}
\end{align}
We now bound $S_T := \sum_{t=1}^T \norm{\nabla F(\x_t) -\gb_t}$ as
\begin{align} \label{eq:init_st_bound}
    S_T \leq \sum_{t=1}^T\rho_1 \sum_{i=1}^{t}(1-\rho_t)^{t-i} a_{i} + \frac{LD}{T} \sum_{t=1}^T \sum_{i=1}^t(1-\rho_t)^{t+1-i}
\end{align}
Using the following lemma from \cite{li2020high}, we can bound the first term on the right hand side
\begin{lemma}{(Li and Orabona 2020 \cite{li2020high}, Lemma 4)}\label{eq:switch}
$\forall T \geq 1$, it holds that
\begin{equation}
    \sum_{t=1}^{T}a_t \sum_{i=1}^t b_i = \sum_{t=1}^T b_t \sum_{i=t}^T a_i
\end{equation}
\end{lemma}
Applying  \cref{eq:switch} (step a) to the first term on the r.h.s. we have
\begin{align}\label{eq:series_A}
    \sum_{t=1}^T\rho_1 \sum_{i=1}^{t}(1-\rho_t)^{t-i} a_{i} &\leq \rho_1 \sum_{t=1}^T (1-\rho_t)^{t}\sum_{i=1}^{t}(1-\rho_t)^{-i} a_{i} \nonumber \\
    &\stackrel{(a)}{=} \rho_1 \sum_{t=1}^T (1-\rho_t)^{-t} a_{t} \sum_{i=t}^T (1-\rho_t)^{i} \nonumber \\
    &= \rho_1 \sum_{t=1}^T a_{t} \sum_{i=t}^T (1-\rho_t)^{i-t}
\end{align}

To get a simple upper bound on this quantity, one could again use the fact that $\rho_t$ is monotonic non-increasing to bound $1-\rho_t \leq  1-\rho_T$. Using the properties of geometric series this would lead to an upper bound of 
\begin{align}
    \rho_1 \sum_{t=1}^T a_{t} \sum_{i=t}^T (1-\rho_t)^{i-t} &\leq \rho_1 \sum_{t=1}^T a_{t} \sum_{i=t}^T (1-\rho_T)^{i-t} \leq \frac{\rho_1}{\rho_T}\sum_{t=1}^T a_{t}
\end{align}
If $\rho_1=\rho_T = \rho$ is a constant, then $\frac{\rho_1}{\rho_T}=1$ and we are just left with a sum of error terms. However, if our momentum term is of the form $\rho_t = \frac{1}{t^\alpha}$, then we have a coefficient $\frac{\rho_1}{\rho_T}=\mathcal{O}(t^\alpha)$ in this bound. In order to achieve a tighter constant bound in the general case of $\rho_t$ we need the following technical lemma:
\begin{lemma}\label{lem:series}
For $\alpha \in (0,1)$ non-inclusive, we have
\begin{equation}
    \sum_{t=1}^\infty (1-\frac{1}{t^\alpha})^t \leq \frac{1}{1-\alpha}\Gamma\left(\frac{1}{1-\alpha}\right)
\end{equation}
Where $\Gamma(z):=\int_0^\infty t^{z-1} e^{-t}dt$ is the Gamma function
\end{lemma}
\begin{proof}(\Cref{lem:series}) Using the well known inequality $(1-\frac{1}{n})^n \leq \frac{1}{e}$ and substituting $n=t^\alpha$ we get
\begin{equation}
    (1-\frac{1}{t^\alpha})^{t^\alpha} \leq e^{-1} \implies (1-\frac{1}{t^\alpha})^{t} \leq e^{-t^{1-\alpha}}
\end{equation}
Where the second inequality can be found by raising both sides to the $t^{1-\alpha}$ power. Since each term in our sequence is positive and decreasing we know that we can bound our series by
\begin{align}
    \sum_{t=1}^\infty (1-\frac{1}{t^\alpha})^t \leq \sum_{t=1}^\infty e^{-t^{1-\alpha}} < \int_0^\infty e^{-t^{1-\alpha}} dt
\end{align}
Lastly we use a change of variables $u = t^{1-\alpha}$, $du = (1-\alpha)t^{-\alpha}dt$ to get the integral into the following form
\begin{align}
    \sum_{t=1}^\infty (1-\frac{1}{t^\alpha})^t &< \frac{1}{1-\alpha}\int_0^\infty e^{-u}u^{\frac{\alpha}{1-\alpha}}du = \frac{1}{1-\alpha}\Gamma(\frac{1}{1-\alpha})
\end{align}
\end{proof}

Using \cref{lem:series} and denoting $K:=\frac{1}{1-\alpha}\Gamma(\frac{1}{1-\alpha})$ as a constant, we can bound the summation in \cref{eq:series_A} by 
\begin{equation}
    \rho_1 \sum_{t=1}^T a_{t} \sum_{i=t}^T (1-\rho_t)^{i-t} \leq \rho_1 K \sum_{t=1}^T a_{t}
\end{equation}

Again using \cref{eq:switch} and \cref{lem:series} we also have an upper bound on the second term in \cref{eq:init_st_bound}: 
\begin{align}
    \frac{LD}{T} \sum_{t=1}^T \sum_{i=1}^t(1-\rho_t)^{t+1-i} \leq LDK
\end{align}
Therefore our bound on $S_T$ simplifies to
\begin{equation}
    S_T \leq \rho_1K\sum_{t=1}^T a_{t} + LDK
\end{equation}
We know by our original assumption that $a_i$ is a zero-mean sub-gaussian random variable, therefore the weighted sum will also be sub-gaussian with variance proxy
\begin{equation}
    \bar{\sigma} ^2 = \rho_1^2K^2\sum_{i=1}^T \sigma_i^2 \leq  K^2\sigma^2T
 \end{equation}
 This means by Hoeffding's inequality
 \begin{equation}
     \mathbb{P}\left( \rho_1 K\sum_{t=1}^T a_{t} \geq \lambda\right) \leq \exp\left\{\frac{-\lambda^2}{2K^2\sigma^2T}\right\}
 \end{equation}
 Rearranging we also find the equivalent statement is true
 \begin{equation}
    \mathbb{P}\left(\sum_{t=1}^T a_{t} \leq  \sqrt{2K^2\sigma^2T\ln(1/\delta)}  \right) \geq 1-\delta
 \end{equation}
 
 Therefore we know that with probability greater than $1-\delta$
 \begin{equation}
     S_T \leq  \sqrt{2K^2\sigma^2T\log(1/\delta)} + LDK
 \end{equation}
 \qedsymbol

%% file: proofs/scg_plus.tex
\subsection{Proof of \Cref{thm:scg_plus}}\label{sec:proof_scg_plus}
Similarly to the proof of \cref{thm:scg_bound}, we first utilize a bound on the variance of the gradient approximation:
\begin{lemma}{(Hassani et al. 2020, Lemma 2)}
Given the gradient approximations $g_t$ as calculated in \cref{alg:scg_plus}, target error $\epsilon$, and batch size $|\mathcal{M}|=1/\epsilon$, we have
\begin{equation*}
    \mathbb{E}\left[\norm{\nabla F(\x_t) - \gh_t}^2\right] \leq (1+\epsilon t) L^2 D^2 \epsilon^2
\end{equation*}
\end{lemma}
Larger batch sizes mean more queries of our stochastic oracle, but have the benefit of reducing the target error $\epsilon$. If we choose $\epsilon = 1/t$ we get the bound:
\begin{equation*}
    \mathbb{E}\left[\norm{\nabla F(\x_t) - \gh_t}^2\right] \leq \frac{2L^2D^2}{t^2}
\end{equation*}
Notice that the residuals between the true gradient and these new estimates decay much more rapidly. Using the Chebyshev inequality we have
\begin{equation}
    \mathbb{P}\left(\norm{\nabla F(\x_t) - \gh_t}^2 \geq \delta^2\frac{2L^2D^2}{t^2}\right) \leq \frac{1}{\delta^2}
\end{equation}
The union bound is likewise
\begin{equation}
    \mathbb{P}\left(\sum_{t=0}^{T-1}\norm{\nabla F(\x_t) - \gh_t}^2 \geq \delta^2\sum_{t=0}^{T-1}\frac{2L^2D^2}{t^2}\right) \leq \frac{T}{\delta^2}
\end{equation}
Now we will begin the proof of \cref{thm:scg_plus}.
\begin{proof}
We intervene at the following inequality found in the original SCG++ expectation bound proof (Hassani et al. 2020, eq. 33):
\begin{equation*}
    F(\x^*)-F(\x_T) \leq (1-\frac{1}{T})^T (F(\x^*)-F(\x_0)) + \sum_{t=0}^{T-1} \frac{1}{2T}\left[\beta_tD^2 + \frac{\norm{\nabla F(\x_t) - \gh_t}^2}{\beta_t}\right] + \frac{LD^2}{2T^2} 
\end{equation*}
This time setting $\beta_t = \frac{\sqrt{2}\delta L}{t}$ we know with probability at least $1-\frac{T}{\delta^2}$
\begin{equation}
        F(\x^*)-F(\x_T) \leq \frac{1}{e} (F(\x^*)-F(\x_0)) + \sum_{t=0}^{T-1} \delta\frac{\sqrt{2}LD^2}{tT} + \frac{LD^2}{2T^2}
\end{equation}
With some rearrangement and simplification
\begin{equation}
    F(\x_T) \geq (1-\frac{1}{e})F(\x^*) - \delta \frac{LD^2}{T} -\frac{LD^2}{2T^2}
\end{equation}
\end{proof}